\documentclass[sigconf, screen, nonacm]{acmart}
\AtBeginDocument{%
  }

\setcopyright{none}
\renewcommand\footnotetextcopyrightpermission[1]{}
\newcommand{\sol}{SPARO}
\usepackage{subcaption}
\usepackage{amsmath} % For math symbols like \pi
\usepackage{amssymb} % For symbols like \curvearrowright
\usepackage{tikz}
\usepackage{quantikz}

\begin{document}

\title{SPARO: Surface-code Pauli-based Architectural Resource Optimization for Fault-tolerant Quantum Computing}

\author{Shuwen Kan}
\email{sk107@fordham.edu}
\affiliation{%
  \institution{Fordham University}
  \country{}  
  % Optionally add \city, \state, \country if desired
}

\author{Zefan Du}
\email{zdu19@fordham.edu}
\affiliation{%
  \institution{Fordham University}
  \country{}  
}

\author{Chenxu Liu}
\email{chenxu.liu@pnnl.gov}
\affiliation{%
  \institution{Pacific Northwest National Laboratory}
  \country{}  
}

\author{Meng Wang}
\email{mengwang@ece.ubc.ca}
\affiliation{%
  \institution{University of British Columbia}
  \country{}  
}

\author{Yufei Ding}
\email{yufeiding@ucsd.edu}
\affiliation{%
  \institution{University of California, San Diego}
  \country{}  
}

\author{Ang Li}
\email{ang.li@pnnl.gov}
\affiliation{%
  \institution{Pacific Northwest National Laboratory}
  \country{}  
}
\affiliation{%
  \institution{University of Washington}
  \country{}  
}

\author{Ying Mao}
\email{ymao41@fordham.edu}
\affiliation{%
  \institution{Fordham University}
  \country{}  
}

\author{Samuel Stein}
\email{samuel.stein@pnnl.gov}
\affiliation{%
  \institution{Pacific Northwest National Laboratory}
  \country{}  
}

% \subtitle{\normalsize{MICRO 2025 Submission
%     \textbf{\#48} -- Confidential Draft -- Do NOT Distribute!!}}
%\author{\normalsize{ISCA 2025 Submission
 %   \textbf{\#NaN} -- Confidential Draft -- Do NOT Distribute!!}}



%%%%%% -- PAPER CONTENT STARTS-- %%%%%%%%

\begin{abstract}

Surface codes represent a leading approach for quantum error correction (QEC), offering a path towards universal fault-tolerant quantum computing (FTQC). However, efficiently implementing algorithms, particularly using Pauli-based computation (PBC) with lattice surgery, necessitates careful resource optimization. Prior work often employs static layouts and simplified error models. These typically fail to capture the full costs and dynamic nature of active computation, leading to resource bottlenecks and suboptimal architectural designs. To address this, we introduce SPARO (Surface-code Pauli-based Architectural Resource Optimization). SPARO features a comprehensive logical error model based on a large corpus of numerical simulations encompassing active Pauli-based computation (PBC) operations—including Pauli product measurements (PPMs), idling qubits, and patch rotations. Our numerical models are integrated within an end-to-end compilation pipeline. SPARO analyzes algorithm-specific bottlenecks arising from constraints such as limited routing areas or magic-state factory throughput. SPARO then dynamically allocates available hardware resources, balancing compute, routing, and magic-state distillation, to minimize space-time overhead and logical error rates for specific workloads. Our simulations demonstrate that SPARO effectively identifies critical resource trade-offs. When evaluated on benchmark circuits, SPARO identifies resource configurations achieving up to 51.11\% logical error rate reductions for 433-qubit ADDER circuits when compared to state-of-the-art static layouts using an identical total resource budget. This dynamic approach enables effective co-optimization of PBC execution and surface-code architectures, significantly improving overall resource efficiency. SPARO will be open sourced.

\end{abstract}
%%e
%\begin{CCSXML}
%<ccs2012>
%  <concept_id>00000000.0000000.0000000</concept_id>
%  <concept_desc>Do Not Use This Code, Generate the Correct Terms for Your Paper</concept_desc>
%  <concept_significance>500</concept_significance>
%  %<concept_id>00000000.00000000.00000000</concept_id>
%  <concept_desc>Do Not Use This Code, Generate the Correct Terms for Your Paper</concept_desc>
%  <concept_significance>300</concept_significance>
%  %<concept_id>00000000.00000000.00000000</concept_id>
%  <concept_desc>Do Not Use This Code, Generate the Correct Terms for Your Paper</concept_desc>
%  <concept_significance>100</concept_significance>
 % <concept_id>00000000.00000000.00000000</concept_id>
%  <concept_desc>Do Not Use This Code, Generate the Correct Terms for Your Paper</concept_desc>
%  <concept_significance>100</concept_significance>
%</ccs2012>
%\end{CCSXML}

%\ccsdesc[500]{Do Not Use This Code~Generate the Correct Terms for Your Paper}
%\ccsdesc[300]{Do Not Use This Code~Generate the Correct Terms for Your Paper}
%\ccsdesc{Do Not Use This Code~Generate the Correct Terms for Your Paper}
%\ccsdesc[100]{Do Not Use This Code~Generate the Correct Terms for Your Paper}

\keywords{Quantum Computing, Fault Tolerance, Surface Code, Quantum Architecture,Quantum Error Correction}

\maketitle

\section{Introduction}

Quantum computers promise a fundamentally new computational paradigm, enabling the efficient solution of certain tasks believed to be intractable for classical supercomputers. To realize this potential, however, physical qubits must be protected from noise that would otherwise corrupt quantum information and preclude reliable computation. Quantum error correction (QEC) offers a promising pathway to mitigating noise, enabling exponential error suppression \cite{roffe2019quantum, beverland2022assessing} provided the underlying hardware error rates fall below a certain threshold~\cite{livingston2022experimental, google2023suppressing, acharya2024quantum, ryan2021realization}. Consequently, QEC is crucial for demonstrating genuine quantum advantage in practical applications.

Among the various QEC codes, the surface code stands out as a leading candidate due to its relatively high noise threshold near $1\%$~\cite{acharya2024quantum, google2023suppressing}, favorable topological layout, and clear path to universal quantum computation using magic-state-based non-Clifford operations~\cite{gidney2021factor,bravyi2005universal}. Techniques such as lattice surgery~\cite{fowler2018low}, where logical qubits are merged or split to implement multi-qubit gates, represent the state-of-the-art for scalable surface-code computations~\cite{horsman2012surface, litinski2019game}. However, efficiently executing algorithms using surface codes demands meticulous resource management. While the basic space and time costs of lattice surgery are well understood~\cite{horsman2012surface, litinski2019game}, how these overheads translate into logical error rates during \emph{active} quantum computation remains largely underexplored. Existing surface-code studies often focus on passive memory benchmarks, and it is not yet fully understood how best to arrange computing, routing, and magic-state resources to minimize logical failures when running arbitrary quantum algorithms.

One promising computational approach to universal fault-tolerant quantum computation (FTQC) within this architecture is Pauli-based computation (PBC)~\cite{PhysRevX.6.021043, litinski2019game}. By re-expressing gates as Pauli product measurements (PPMs) and rotations, PBC seamlessly integrates with lattice surgery and allows Clifford gates to be efficiently commuted out of the circuit \cite{PhysRevX.6.021043}. However, this simplification often increases the complexity of non-Clifford operations and measurements, resulting in high-weight PPMs that require intricate ancilla qubit routing and logical qubit rotations. As we show, the errors associated with these complex ancilla paths can quickly dominate the overall logical failure rate. Therefore, optimizing FTQC performance with PBC necessitates models that accurately capture the error characteristics and dynamic resource demands during active computation. Prior architectural models often lack the granularity to capture the full costs and dynamic interplay between: (1) specific PBC operations (complex PPMs, Y-basis measurements, patch rotations), (2) physical resource constraints (limited routing area affecting PPM ancilla paths, finite magic-state factory throughput inducing latency), and (3) the algorithm's evolving computational patterns. Quantifying PBC’s full resource requirements and the resulting logical failure rate trade-offs is non-trivial. Addressing these open questions requires a comprehensive model that models how each operation contributes to logical errors and how different architectural choices (e.g., routing area vs. number of magic-state factories) impact performance under resource constraints. This incomplete picture hinders the development of resource allocation and compilation strategies that can adapt to dynamic needs and minimize logical errors effectively, especially when considering flexible allocation of additional resources beyond the bare minimum.

To address these challenges, we introduce SPARO (Surface-code Pauli-based Architectural Resource Optimization), a dynamic resource-allocation framework for mapping transpiled PBC circuits onto a surface-code architecture. SPARO integrates a comprehensive logical error model, meticulously simulated and validated, that accounts not only for the multiple failure modes of multi-qubit PPMs (critically considering ancilla path length) but also for the cumulative effects of idle qubits, boundary reorientations (patch rotations), and other ancilla-based operations. This model enables an end-to-end optimization pipeline that translates high-level quantum algorithms into scheduled surface-code lattice-surgery operations. Crucially, SPARO analyzes the execution trace to identify algorithm-specific bottlenecks (e.g., contention for routing resources, delays awaiting T-states) and then strategically allocates available hardware resources, thus balancing the needs of computation, routing, and magic-state distillation to reduce the space-time overhead and lower logical error rates for the target workload.

Specifically, SPARO contributes:
\begin{itemize}
    \item \textbf{Comprehensive Error Modeling.} We perform extensive simulations of Pauli product measurements on surface codes at different distances and physical error rates, revealing that ancilla-path length is often the primary factor in logical failures. This leads to an algorithm-level noise model that translates hardware overheads into logical failure probabilities, providing the quantitative foundation needed for fine-grained, dynamic resource optimization.

    \item \textbf{End-to-End Compilation and Optimization Pipeline.} SPARO compiles high-level quantum algorithms into scheduled lattice-surgery operations to a surface-code topology, employing a suite of optimization techniques that leverage deep insights into architectural constraints for large-scale, fault-tolerant computation, including Steiner-tree based PPM routing and measurement scheduling.

    \item \textbf{Bottleneck Analysis and Dynamic Allocation.} By examining how limited resources (e.g., routing regions, magic-state factories) impact overall logical error rates and runtime, SPARO highlights critical bottlenecks. We address these issues via dynamic resource distribution, enabling more efficient workload management and improved algorithm-level reliability, demonstrably achieving lower logical errors compared to static layouts using the same total resource footprint.
\end{itemize}
\section{Background}

\subsection{Quantum Error Correction}
Achieving fault-tolerant quantum computing remains a significant engineering challenge. In many current hardware platforms~\cite{smith2022scaling}, physical error rates exceed the thresholds required for reliable computation, and topological constraints limit connectivity. Consequently, quantum error correction (QEC) is widely regarded as the cornerstone for achieving quantum advantage at scale. QEC employs quantum error-correcting codes to encode logical qubits within a larger physical Hilbert space. By performing syndrome measurements and utilizing classical decoders, errors on the physical qubits can be detected and corrected. When physical error rates are below a code-specific threshold, QEC provides an exponential suppression of the logical error rate as the code distance increases~\cite{livingston2022experimental,google2023suppressing,acharya2024quantum,ryan2021realization}.

\subsubsection{Surface Code}
The surface code~\cite{fowler2012surface} is among the most extensively investigated QEC codes, primarily due to several advantages. First, it only requires local couplings on a two-dimensional grid of qubits with only nearest-neighbor interactions, which is well-suited to leading hardware platforms such as superconducting systems. Second, it exhibits a high error threshold, tolerating physical error rates approaching $1\%$~\cite{acharya2024quantum,google2023suppressing}. Third, it offers a clear pathway to universal fault-tolerant quantum computation (FTQC) through the Clifford+T gate set or Pauli-based computation~\cite{gidney2021factor,fowler2018low,litinski2019game,bravyi2005universal}. Within the surface code, Clifford gates can be implemented via transversal gate operations, code deformation or lattice surgery, while the resource-intensive non-Clifford T gates ($\ket{T}=\frac{1}{\sqrt{2}}(\lvert 0 \rangle + e^{i\pi/4}\,\lvert 1 \rangle)$) are typically realized using ancillary magic states prepared via distillation or cultivation \cite{gidney2024magic,litinski2019magic}.

\subsection{Pauli based Pauli-based computation}

\begin{figure}
    \centering
    \includegraphics[width=1\linewidth]{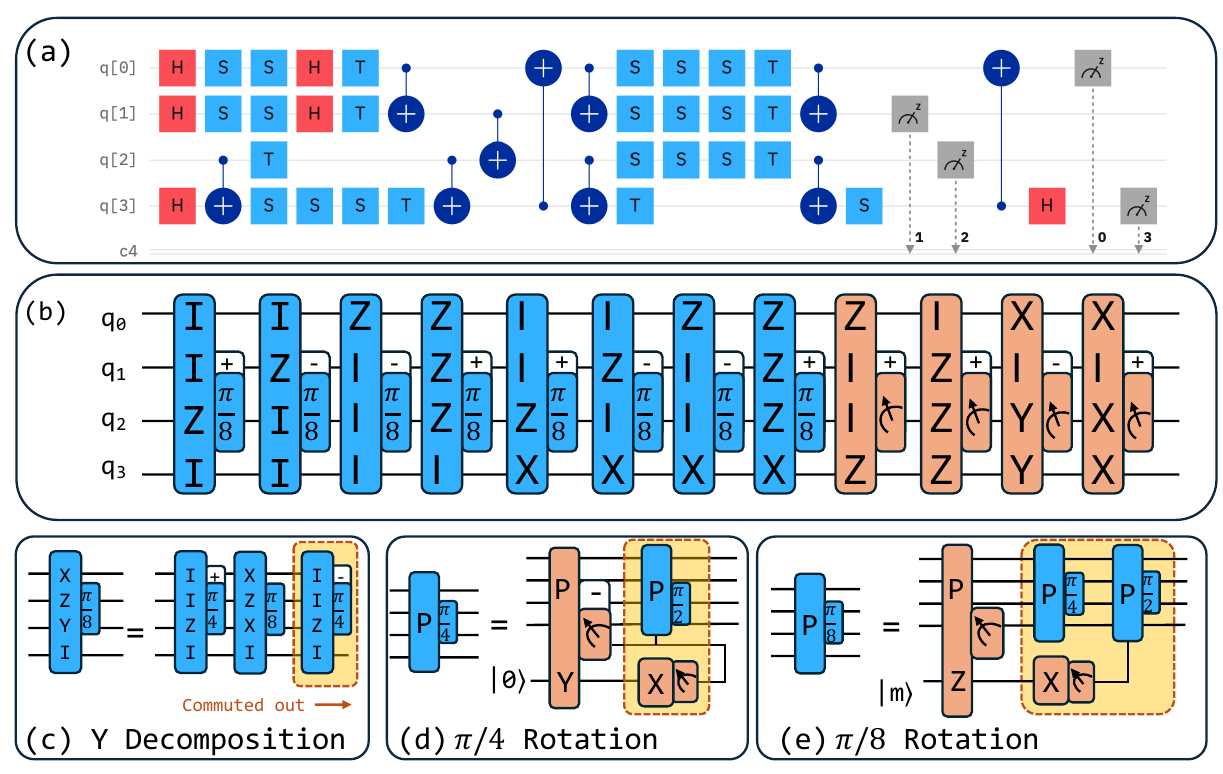}
    \caption{\textbf{A 4-qubit Adder circuit: (a) Clifford+T circuit (b) equivalent PBC compute model for this circuit composed of a sequence of Pauli product rotations and measurements with alternating signs. (c), (d) and (e) illustrate the rules to effectively translate Pauli rotations to Pauli product measurements. Note that the framed $\pi/4$ and $\pi/2$ rotation can be commuted out of the circuit}}
    \label{fig:enter-label}
\end{figure}

Pauli-Based Computation (PBC) is a universal quantum computing model that leverages Pauli product measurements (PPMs) and classical processing to execute circuits~\cite{PhysRevX.6.021043}. It is a natural extension of the Clifford + T computational model. In the context of FTQC, PBC typically involves transpiling arbitrary quantum circuits into the Clifford+T gate set~\cite{litinski2019game}, which includes Hadamard ($H$), phase ($S$), $\pi/8$ ($T$), and CNOT ($\text{CX}$) gates.

Each gate can be expressed as a sequence of Pauli rotations $e^{i\theta P}$, where $P \in \{I, X, Y, Z\}$ is a Pauli operator and $\theta$ is the rotation angle. Standard PBC decompositions include:
\( H = Z_{\pi/4} X_{\pi/4} Z_{\pi/4} \),
\( S = Z_{\pi/4} \),
\( T = Z_{\pi/8} \), and
\( \text{CX} = ZX_{\pi/4} \cdot ZI_{-\pi/4} \cdot IX_{-\pi/4} \).
Clifford gates correspond to $\pi/4$ rotations, while the non-Clifford T gate corresponds to a $\pi/8$ rotation. A key advantage of PBC is that Clifford operations ($e^{i\frac{\pi}{4} P}$) can be commuted out of the circuit. This commutation follows specific rules: if a Clifford rotation $P$ commutes with a non-Clifford rotation $P'$, they can be reordered freely; if they anti-commute ($PP' = -P'P$), commuting the Clifford past the non-Clifford modifies the non-Clifford operator ($P' \rightarrow iPP'$). Commuting Cliffords past measurements simplifies the quantum operations required, resulting in only non-Clifford Pauli product rotations and measurements, reducing circuit depth and complexity~\cite{PhysRevX.6.021043, litinski2019game}. However, Pauli based computation trades off completely removing Clifford operators for higher weight operators, requiring more complex and erroneous ancilla routing. 

\subsubsection{Pauli Product Measurement (PPM)}
A Pauli Product Measurement measures the eigenvalue of a product of Pauli operators acting on multiple qubits (e.g., $X_1 Z_2 X_3$). In surface code architectures, PPMs are typically implemented using lattice surgery, where an ancilla patch is prepared, interacts with the relevant boundaries of the data qubit patches, and is then measured~\cite{fowler2018low, litinski2019game}. As mentioned, PBC often leads to high-weight PPMs (involving many qubits). Implementing these requires correspondingly complex ancilla patches and interaction paths, which, as we will show, can become significant sources of logical error.

\subsubsection{Magic State}
Realizing non-Clifford operations, such as the T gate, fault-tolerantly requires high-fidelity ancillary magic states (specifically, $\ket{T}$ states). Since directly preparing such states is error-prone, protocols like magic state distillation~\cite{bravyi2012magic, litinski2019magic} and magic state cultivation~\cite{gidney2024magic} are employed.
\begin{itemize}
    \item \textbf{Magic State Distillation:} Iteratively combines multiple noisy magic states to probabilistically produce fewer, higher-fidelity magic states. This incurs significant resource overhead in both space (multiple input states needed) and time (multiple rounds may be required).
    \item \textbf{Magic State Cultivation:} Injects a state into a small error-correcting code (like a color code patch) and incrementally increases its effective code distance through checks. While potentially offering lower overheads than distillation, cultivation protocols can be complex and sensitive to the underlying physical error rate.
\end{itemize}
Both methods introduce latency and consume resources, creating potential bottlenecks if magic state production cannot keep pace with algorithmic demand.

\subsection{Fault-Tolerant Architecture}
Executing algorithms fault-tolerantly requires mapping logical operations onto the physical architecture while managing resource overheads. While early proposals involved braiding defects within the surface code, this approach often incurs substantial spatial overhead to maintain equivalent logical error suppression when compared to lattice surgery. As such, lattice surgery~\cite{fowler2018low} has emerged as the preferred method for large-scale computation due to its reliance on geometrically local operations.

\subsubsection{Lattice Surgery}
Lattice surgery operates on rotated surface code patches~\cite{fowler2018low}. Logical operations, like multi-qubit measurements (PPMs) or controlled gates, are performed by merging and splitting logical qubit patches using adjacent ancilla regions. This involves temporarily enabling specific stabilizer measurements along the boundary between data qubit patches and ancilla patches. For example, measuring a Pauli product like $X_1 X_2$ can be achieved by preparing an ancilla patch, performing boundary measurements corresponding to X between the ancilla and qubit 1, and between the ancilla and qubit 2, followed by measuring the ancilla~\cite{litinski2019game}. 

\subsubsection{Tile-Based Computation} % Corrected title
Surface code computations are naturally viewed in terms of computational "tiles"~\cite{litinski2019game, horsman2012surface}. Each tile represents a surface code patch encoding one logical qubit, typically with a fixed physical size determined by the desired code distance $d$. Logical $X$ and $Z$ operators are associated with the boundaries of these tiles, and is visualized in Figures \ref{fig:layout} and \ref{fig:rotation}. Boundaries must be oriented according to the computation.  Operations like lattice surgery involve interactions between adjacent tiles (data or ancilla patches). To maintain the overall code distance during these interactions, ancilla patches used for surgery must also be constructed with a matching code distance. This tile-based perspective facilitates modular design and resource estimation for complex algorithms.

\section{Challenges and Opportunities}

While surface codes and techniques like lattice surgery provide a path to FTQC, significant challenges remain in optimizing resource usage and minimizing logical errors during the execution of complex algorithms. Static approaches and incomplete models prohibit optimizations and accurate modeling.

\subsection{Challenge: Modeling Logical Errors in Active Surface-Code Computation}

Existing analyses of surface code performance predominantly rely on passive memory experiments~\cite{acharya2024quantum}. While crucial for demonstrating exponential error suppression with increasing code distance, these benchmarks measure state preservation in the absence of operations and thus neglect the complexities inherent in active computation. Consequently, they fail to capture the full spectrum of error sources encountered during algorithm execution.

\begin{figure}
    \centering
    \includegraphics[width=1\linewidth]{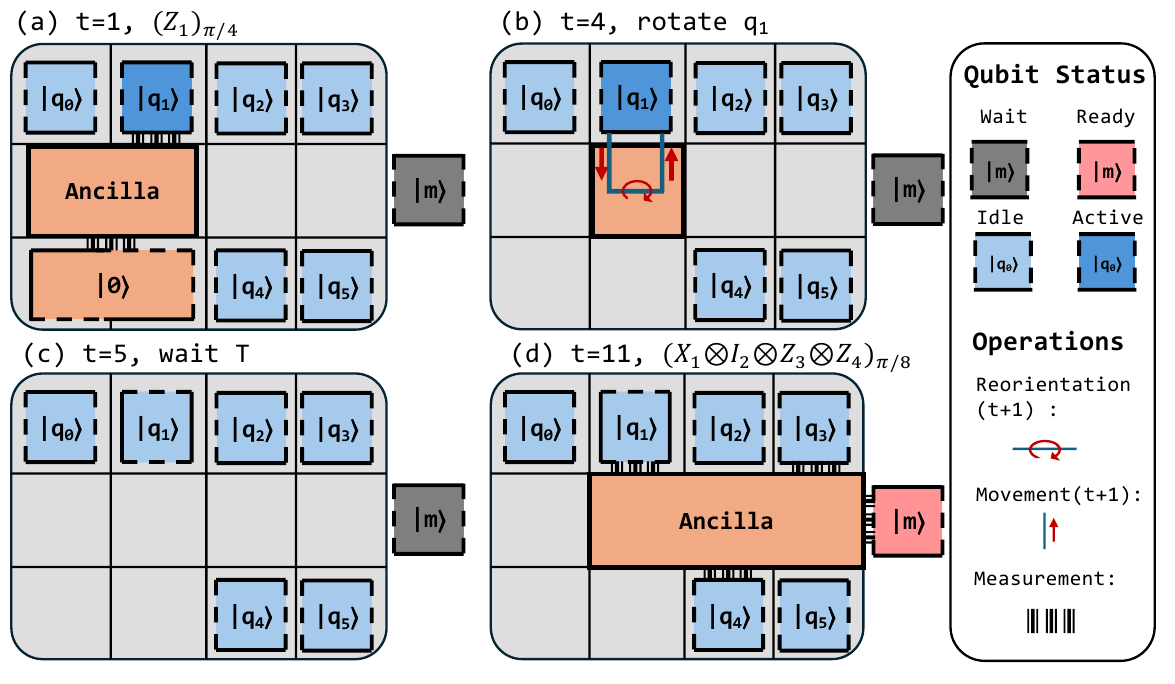}
    \caption{Illustration of operational overhead from logical qubit patch manipulation (e.g., rotation for boundary realignment). (a) Initial state of a section of the layout at time t=0, including logical qubits $|q_0\rangle...|q_7\rangle$, a magic state port $|m\rangle$, and routing space. (b) State after 4 time steps (t=4). Qubit $|q_1\rangle$ has undergone active operations (indicated by legend symbols for reorientation/movement), utilizing the routing region (orange tiles are active) and potentially an ancilla state $|0\rangle$. Such operations are necessary for aligning boundaries for certain gates or measurements in PBC and contribute significantly to the overall execution time and potential error accumulation, highlighting a key overhead managed by SPARO.}
    \label{fig:rotation}
\end{figure}

Pauli-based computation (PBC)~\cite{PhysRevX.6.021043, litinski2019game}, although simplifying Clifford gate handling, introduces specific error modeling challenges. A common outcome of PBC is the need for high-weight Pauli product measurements (PPMs). Implementing these via lattice surgery necessitates complex routing paths for ancilla qubits. As our simulations demonstrate, the ancilla path length during PPMs is a dominant factor contributing to logical errors, an effect often underestimated or overlooked in simplified models. Furthermore, other operations intrinsic to active computation, such as logical qubit rotations (boundary reorientations necessary for gate alignment, see Figure~\ref{fig:rotation}) and periods where qubits idle while awaiting resources (e.g., T-states from magic state factories), cumulatively contribute to the final logical failure probability. Achieving accurate performance prediction and enabling effective optimization requires a comprehensive error model that identifies and quantifies \emph{all} significant error sources during active computation and understands their dependence on architectural choices and operational parameters.

\subsection{Challenge: Resource Management and Time Overheads in Fixed Topologies}
\begin{figure}
    \centering
    \includegraphics[width=0.95\linewidth]{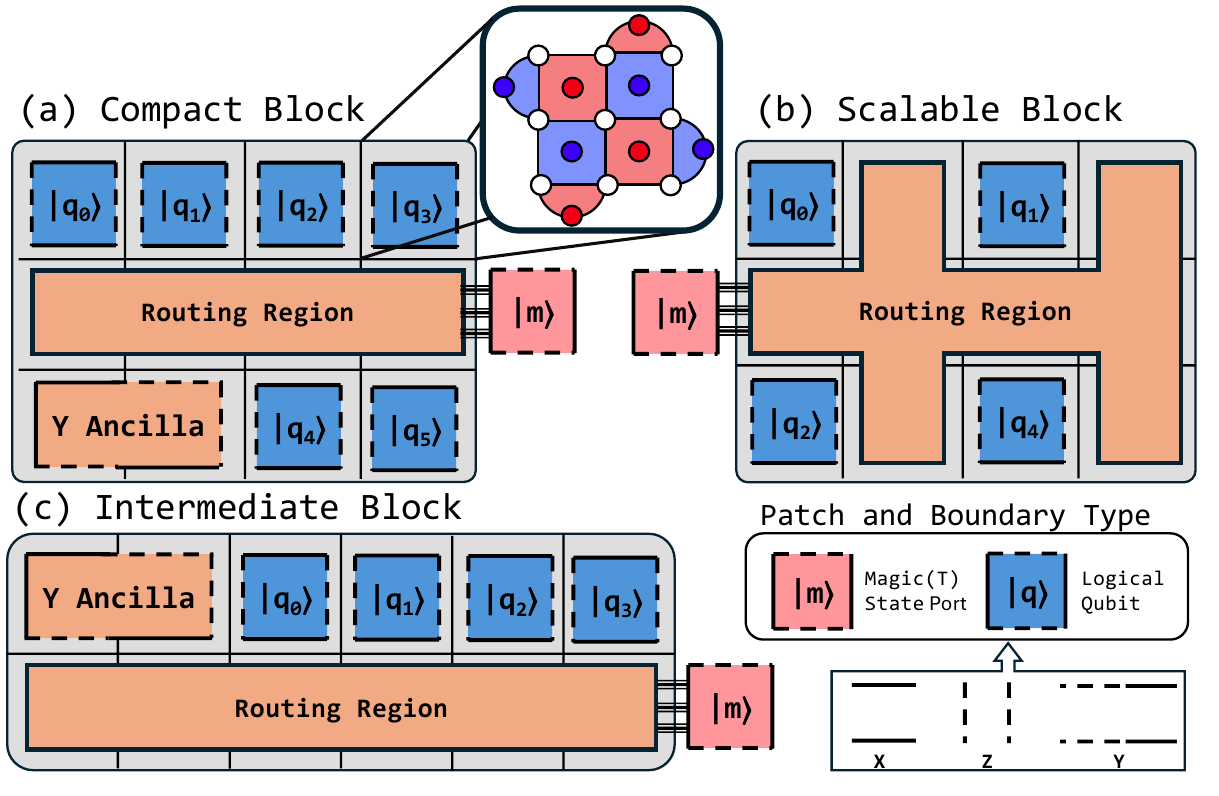}
    \caption{ Examples of static surface code layouts based on prior work. (a) Litinski's 'compact block'~\cite{litinski2019game}, prioritizing qubit density. (b) A tile-based layout with a fixed compute-to-routing ratio, similar to Horsman et al.~\cite{horsman2012surface}. (c) Litinski's 'intermediate block'~\cite{litinski2019game}, attempting to balance compute and routing space. Each logical qubit tile (blue squares) has distinct X and Z boundaries (indicated in the legend). Gray areas are dedicated routing regions for ancilla paths during lattice surgery. These layouts represent fixed allocations of hardware resources and do not explicitly incorporate magic state factory positioning or allow for dynamic resizing based on algorithmic needs. }
    \label{fig:layout}
\end{figure}

Current strategies for surface code based PBC often employ static, fixed-topology designs, where hardware resources are partitioned a priori into fixed regions for computation, routing, and magic-state preparation~\cite{fowler2018low, litinski2019game, horsman2012surface}. Examples include layouts with predetermined compute-to-routing area ratios~\cite{horsman2012surface} or standardized "compact," "intermediate," and "fast" block architectures~\cite{litinski2019game} (visualized in Figure~\ref{fig:layout}). While providing a basis for design, these static approaches suffer from inherent inflexibility, potentially leading to substantial space-time overheads and under-utilization.

A critical bottleneck arises from provisioning magic-state factories. Protocols like distillation are necessary for non-Clifford T gates but are typically much slower than gate execution; for instance, a common 15-to-1 distillation protocol requires 11 cycles to produce a T-state that is consumed in a single cycle~\cite{litinski2019magic}. If the number of factories is fixed based on average demand, it may become a bottleneck during algorithm phases with high T-gate density ("T-bursts"), forcing compute qubits to idle and accumulate errors~\cite{beverland2022assessing}. Conversely, factories may be underutilized during qubit reorientation phases.

This inflexibility highlights a fundamental space-time trade-off between routing capacity, compute density, and magic-state production rate. Compact layouts minimize area but may serialize operations or require costly patch manipulations due to limited routing. Layouts with abundant routing (like the "fast" block) reduce pathfinding overheads but consume significant space, potentially limiting computational density or factory count. Static designs commit to a single point in this trade-off space, which is unlikely to be optimal across diverse algorithms or even different execution phases of one algorithm. The impact of PPM ancilla length on errors further complicates this, as routing limitations directly translate to higher failure rates.

\subsection{Opportunity: Integrated, Algorithm-Aware Resource Management}

The limitations of static layouts designed around worst-case assumptions highlight the need for more sophisticated approaches. Relying on metrics derived from static scenarios (e.g., the maximum time for any possible PPM ~\cite{litinski2019game}) overlooks the variable nature of a PBC quantum algorithm. The distribution of T-gates and the complexity of PPMs often vary significantly throughout an algorithm's execution.

This variability presents a substantial opportunity for optimization through dynamic, algorithm-aware resource management. By analyzing algorithm-specific patterns—such as T-gate bursts, multi-qubit interaction densities (high-weight PPMs), and measurement schedules—hardware resource allocation and operation scheduling can be tailored to meet instantaneous demands. For instance, routing resources could be temporarily expanded during complex PPM phases, or magic state production could be ramped up preceding T-bursts.

This perspective motivates the co-design philosophy embodied by SPARO. It integrates architecture-level constraints with algorithm scheduling, informed by detailed, simulation-driven error quantification (capturing contributions from PPMs, rotations, idling, etc.). Such integration enables dynamic decisions, like adjusting the effective allocation between routing and factories or even modifying local code distances if feasible, to mitigate the impact of bottlenecks and high-weight operations.

Ultimately, a holistic view that jointly considers hardware topology, realistic error models, and algorithm-specific computational patterns is essential for efficient FTQC. Moving beyond rigid, worst-case planning to adaptive resource management allows architectures to leverage algorithmic structure, reduce idle overhead, and achieve lower overall logical error rates. This adaptive strategy is key to unlocking the full potential of surface code computation.
\section{\sol{}}

\begin{figure*}
    \centering
    \includegraphics[width=1\linewidth]{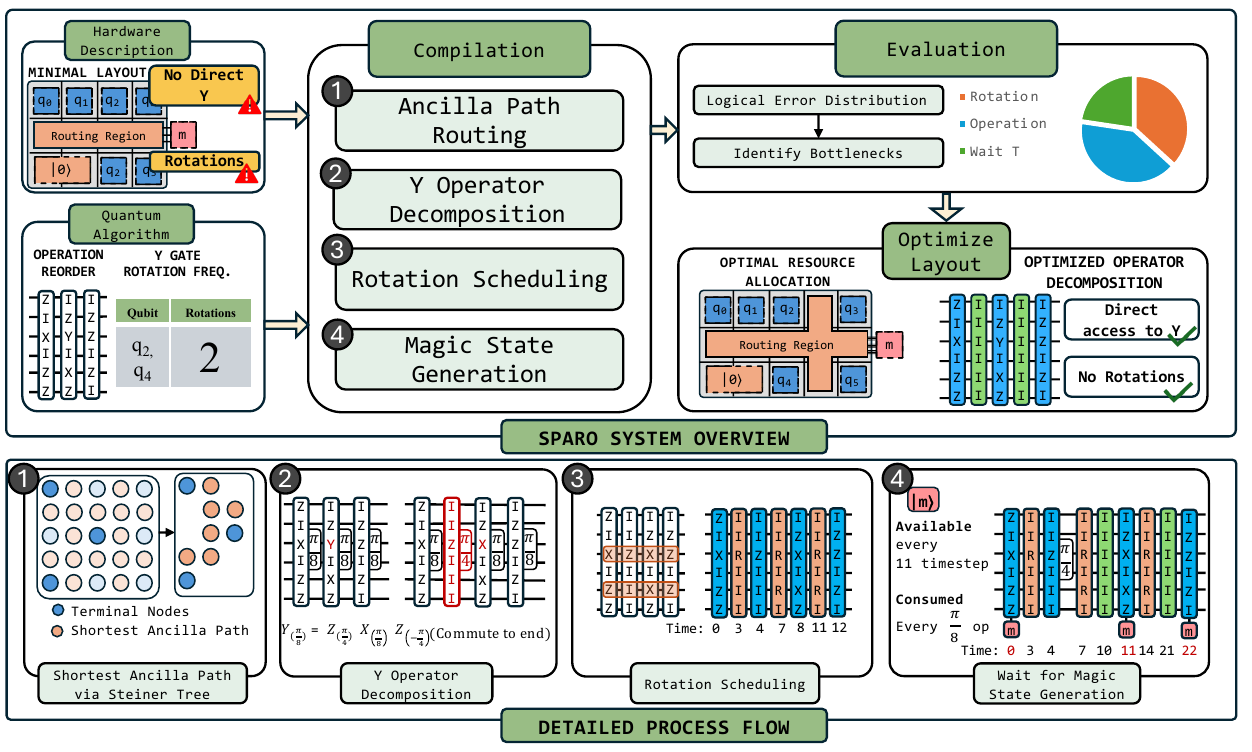}
    \caption{Overview of the SPARO framework. SPARO takes hardware resource constraints (defining an initial minimal layout) and a target quantum algorithm (analyzed for qubit interaction patterns and rotation frequencies) as input. Its compilation pipeline optimizes the algorithm's execution on a surface code architecture by employing techniques such as shortest ancilla pathfinding (Steiner trees) for PPMs, T-gate reordering, measurement parallelization, and optimized Y-operator decomposition, while accounting for magic state generation latency. SPARO analyzes the resulting execution schedule and resource demands to predict the logical error distribution, identifying bottlenecks associated with PPM operations (Op), patch rotations/realignments (Rotation), and waiting for T-states (Wait T). Based on this analysis, SPARO determines a dynamic allocation of the available hardware resources (e.g., adjusting routing space vs. T-factory count) to generate an optimized layout and schedule aimed at minimizing the overall logical error rate and space-time overhead for the specific algorithm.}
    \label{fig:system}
\end{figure*}

This section presents our solution framework, SPARO, designed for optimizing Pauli-based computations (PBC) on surface-code architectures. As illustrated in Figure~\ref{fig:system}, SPARO integrates detailed error modeling with algorithm-aware compilation and dynamic resource allocation. We begin by outlining our comprehensive logical error model (\S\ref{subsec:error_model}), followed by the definition of a minimal architectural layout used as a baseline (\S\ref{subsec:min_arch_layout}). We then describe SPARO's transpilation pipeline (\S\ref{subsec:transpilation}) and detail how it performs dynamic hardware resource allocation based on bottleneck analysis (\S\ref{subsec:dynamic_allocation}). Finally, we explain how SPARO generalizes to alternative magic-state preparation protocols (\S\ref{sec:magic_state_cultivation}).

\subsection{Developing a Comprehensive Logical Error Model}
\label{subsec:error_model}

To accurately guide resource allocation, SPARO relies on a detailed logical error model derived from simulations of primitive operations in PBC on a surface code architecture. The model accounts for errors arising from:
\begin{enumerate}
    \item \textbf{Pauli Product Measurements (PPMs)}: Multi-qubit measurements forming the core of PBC. The error significantly depends on the ancilla path length required.
    \item \textbf{Y-Basis Measurements}: Often implemented via twist defects~\cite{PhysRevResearch.4.023090}, which can have higher error rates than standard X/Z measurements due to complex stabilizer interactions and require dedicated space, potentially making them unavailable in minimal layouts.
    \item \textbf{Idling Operations}: Logical qubits accumulate errors even when not actively participating in computation during a given time step.
    \item \textbf{Surface-Code Patch Rotations}: Operations to realign logical boundaries (e.g., for specific gate implementations) or move patches introduce errors, potentially including moving back to the original position if layout constraints require it.
\end{enumerate}

\paragraph{Layer-Wise Logical Error Calculation.}
We analyze errors on a layer-by-layer basis, corresponding to the time steps in the scheduled computation. In layer \(t\), let \(P_{\mathrm{op}}^{(t)}\) be the probability of failure due to active operations (PPMs, Y-measurements, rotations), which depends on the number of qubits involved \(n_{q}\) and specifics like ancilla path length \(\ell_{\mathrm{anc}}\) for PPMs, i.e., \(P_{\mathrm{op}}^{(t)} = f(n_{q}, \ell_{\mathrm{anc}}, \text{op\_type})\). Let \(P_{\mathrm{idle}}^{(t)}\) be the probability that at least one of the \(n_{\mathrm{idle}}\) idling qubits fails during this layer, calculated as \(P_{\mathrm{idle}}^{(t)} = 1 - \prod_{i=1}^{n_{\mathrm{idle}}} (1 - p_{i,t})\), where \(p_{i,t}\) is the idle error probability for qubit \(i\) in layer \(t\).

The total logical error probability for layer \(t\), \(p_L^{(t)}\), assuming operation and idle errors are independent primary failure modes within the layer, is the probability that either an operation fails or an idle qubit fails:
\( p_L^{(t)} = 1 - (1 - P_{\mathrm{op}}^{(t)})(1 - P_{\mathrm{idle}}^{(t)}) \).
The total logical error probability over \(T\) layers is then given by the union bound across layers (assuming independence):
\begin{equation}
    p_L^{\text{total}} = 1 - \prod_{t=1}^{T} (1 - p_L^{(t)}) \approx \sum_{t=1}^{T} p_L^{(t)} \quad \text{for } p_L^{(t)} \ll 1.
    \label{eq:total_logical_error}
\end{equation}
This model highlights how errors accumulate over the computation's duration (\(T\)). The function \(f\) quantifying \(P_{\mathrm{op}}^{(t)}\) is derived from detailed simulations, capturing the crucial dependence on factors like \(\ell_{\mathrm{anc}}\).

\paragraph{Implications for \(Y\)-Measurements.}
This layered model allows incorporating the specific costs of different operations. Y-measurements implemented via twist defects~\cite{PhysRevResearch.4.023090}, for example, may contribute a higher \(p_L^{(t)}\) compared to standard PPMs due to their sensitivity to errors, especially if multiple Y-measurements occur concurrently. SPARO accounts for these implementation-specific costs when available resources permit such operations.

\subsection{Minimal Architectural Layout Baseline}
\label{subsec:min_arch_layout}

To establish a baseline and analyze inherent overheads, we define a minimal layout using the fewest resources necessary to execute a generic PBC algorithm. This layout typically includes:
\begin{itemize}
    \item A compute region sized for the required $n$ logical qubits.
    \item A minimal routing region capable of connecting necessary qubits.
    \item A single magic-state factory (e.g., 11 tiles for a 15-to-1 distillation protocol~\cite{litinski2019magic}).
    \item Ancillary resources (e.g., 2 tiles) potentially needed for Y-measurement decomposition if twist defects are not used.
\end{itemize}
This configuration is inspired by compact block designs (e.g., Figure~\ref{fig:layout}(a)). While area-efficient, it inherently suffers from temporal overheads due to resource contention in the form of:
\begin{enumerate}
    \item \textbf{Magic-State Latency}: T-gate operations must wait for the single factory to produce a $|T\rangle$ state.
    \item \textbf{Routing/Rotation Delays}: Limited routing space may serialize PPMs or necessitate time-consuming patch rotations for boundary alignment.
\end{enumerate}
This minimal layout serves as the starting point from which SPARO considers adding resources dynamically.

\subsection{SPARO Transpilation Pipeline}
\label{subsec:transpilation}

Given a PBC circuit, SPARO employs a compilation pipeline (Figure~\ref{fig:system}) to generate an optimized schedule for a target surface code architecture. Key stages include:

\subsubsection{PPM Ancilla Pathfinding via Steiner Trees}
Minimizing ancilla path length (\(\ell_{\mathrm{anc}}\)) for PPMs is critical for reducing logical errors. Since data qubits act only as leaves, routing occurs solely through available ancilla patches in the routing region. SPARO models this as a Steiner Tree problem \cite{hwang1992steiner} on the graph of available ancilla patches, where terminal nodes are ancillas adjacent to the required data qubit boundaries. To manage complexity, a heuristic prunes candidate Steiner vertices based on Manhattan distance to the terminals. A classical Steiner Tree solver then finds the minimum-cost ancilla path, minimizing \(\ell_{\mathrm{anc}}\) and thus the PPM error contribution.

\subsubsection{Operation Scheduling}
SPARO optimizes the schedule to minimize latency and overhead:
\begin{itemize}
    \item \textbf{Rotation Scheduling:} While T-gates across layers are fixed, operations within a layer can be reordered. SPARO minimizes basis shift overhead (e.g., for Y-measurements) by optimizing the intra-layer order, scheduling operations such that we minimize rotation calls.
    \item \textbf{Measurement Parallelization:} Independent measurements (acting on disjoint qubit sets) can run in parallel. SPARO constructs a measurement interaction graph and uses graph coloring to find the minimum number of sequential measurement steps, maximizing parallelism.
\end{itemize}

\subsubsection{Qubit Mapping}
To minimize communication distances (and thus \(\ell_{\mathrm{anc}}\) and rotation overheads), SPARO maps logical qubits to physical tile coordinates using a two-stage process:
\begin{enumerate}
    \item \textbf{Greedy Initialization:} Places frequently used qubits (based on T-layer analysis) near the magic-state factories.
    \item \textbf{Simulated Annealing Refinement:} Iteratively swaps qubit positions to minimize a cost function based on total Manhattan distance for interacting qubits in the algorithm.
\end{enumerate}
\subsection{Dynamic Resource Allocation Strategy}
\label{subsec:dynamic_allocation}

A core contribution of SPARO is its ability to dynamically allocate hardware resources beyond the minimal baseline layout (\S\ref{subsec:min_arch_layout}) to minimize the overall logical error rate (\(p_L^{\text{total}}\)) for a specific target algorithm, given a total resource budget. This involves strategically distributing additional tiles between expanding the routing region (\(R\)) and increasing the number of T-state factories (\(N_F\)). The process integrates insights from algorithm analysis, compilation, and the logical error model, as illustrated in Figure~\ref{fig:system}.

\subsubsection{Identifying Resource Demands via Algorithm and Pipeline Analysis}
\label{subsubsec:pattern_analysis} % Renamed for clarity

The optimal resource allocation heavily depends on the algorithm's computational structure and the overheads revealed during compilation. SPARO first analyzes the input PBC circuit and the output of its compilation pipeline (\S\ref{subsec:transpilation}) to quantify demands:

\begin{itemize}
    \item \textbf{T-State Demand:} High frequency of T-gates, particularly concentrated bursts, indicates a high demand for T-state generation. The pipeline explicitly calculates the time qubits spend idling while waiting for T-states (\(T_{WaitT}\)).
    \item \textbf{Routing Demand (from PPMs):} Frequent execution of high-weight PPMs, especially those involving distant qubits (identified during qubit mapping and Steiner tree pathfinding), results in long ancilla paths (\(\ell_{\mathrm{anc}}\)). This increases both the execution time (\(T_{Op}\)) and the operational error probability (\(P_{\mathrm{op}}\)) for PPMs.
    \item \textbf{Routing Demand (from Rotations):} Frequent requirement for logical qubit rotations (e.g., for basis changes like Z to Y) incurs significant time overhead (\(T_{Rotation}\)) due to patch manipulations, especially in constrained routing areas.
\end{itemize}
These factors are quantified during the initial compilation run using a baseline configuration (e.g., the minimal layout).

\subsubsection{Bottleneck Identification and Marginal Resource Analysis}
\label{subsubsec:marginal_analysis} % Renamed for clarity

Using the data gathered above, SPARO identifies the dominant performance bottleneck and estimates the benefit of adding resources incrementally.

\begin{enumerate}
    \item \textbf{Quantify Bottlenecks:} After the initial compilation, SPARO aggregates the total time spent and the total predicted logical error contributed by each category across all layers of the computation:
        \begin{itemize}
            \item \(T_{WaitT}^{total}, E_{WaitT}^{total}\): Total time and predicted error from idling for T-states.
            \item \(T_{Rotation}^{total}, E_{Rotation}^{total}\): Total time and predicted error from patch rotations/realignments.
            \item \(T_{Op}^{total}, E_{Op}^{total}\): Total time and predicted error from active PPM operations (sensitive to \(\ell_{\mathrm{anc}}\)).
        \end{itemize}
        The primary bottleneck is identified as the category contributing most significantly to the total execution time (\(T_{total} = T_{WaitT}^{total} + T_{Rotation}^{total} + T_{Op}^{total}\)) or, more critically, the total logical error (\(p_L^{\text{total}} \approx E_{WaitT}^{total} + E_{Rotation}^{total} + E_{Op}^{total}\)).

    \item \textbf{Estimate Marginal Gain:} SPARO employs a heuristic-based marginal analysis to decide where to allocate the next unit of available resource area (e.g., a fixed number of tiles). It estimates the potential reduction in \(p_L^{\text{total}}\) from adding resources to either factories or routing:
        \begin{itemize}
            \item \textbf{Gain from adding T-Factory resources} (costing \(Area_{factory}\) tiles):
                If \(E_{WaitT}^{total}\) is dominant, adding factories is prioritized. SPARO estimates the reduction in \(T_{WaitT}^{total}\) (e.g., assuming faster average T-state availability) and recalculates the corresponding reduction in \(E_{WaitT}^{total}\) using the error model. The impact on other error terms is assumed small for this estimation step.
            \item \textbf{Gain from adding Routing resources} (costing \(Area_{routing\_unit}\) tiles):
                If \(E_{Op}^{total}\) or \(E_{Rotation}^{total}\) dominate, adding routing space is prioritized. SPARO estimates the reduction in the average \(\ell_{\mathrm{anc}}\) based on increased routing availability (potentially using geometric arguments or lookup tables from pre-simulation) and recalculates the reduction in \(E_{Op}^{total}\). It similarly estimates the reduction in rotation delays (\(T_{Rotation}^{total}\)) and its error contribution (\(E_{Rotation}^{total}\)). The impact on \(E_{WaitT}^{total}\) is assumed small.
        \end{itemize}

        The estimated benefit is quantified as the reduction in total logical error per resource unit added. SPARO compares the estimated marginal error reduction from adding factory resources (\(\frac{|\Delta E_{WaitT}^{total}|}{Area_{factory}}\)) versus adding routing resources (\(\frac{|\Delta E_{Op}^{total} + \Delta E_{Rotation}^{total}|}{Area_{routing\_unit}}\)) and prioritizes the resource type yielding the larger estimated reduction in \(p_L^{\text{total}}\).

    \item \textbf{Iterative Allocation:} SPARO greedily allocates the next unit of resource area to the option (factories or routing) offering the best estimated marginal benefit (\(\max(\Delta p_L^{\text{total}} / Area_{resource})\)). The layout configuration (\(N_F, R\)) is updated. The compilation pipeline (at least the affected parts like pathfinding or scheduling) is re-run with the new configuration to get updated bottleneck statistics (\(T^{total}\), \(E^{total}\) components). This iterative process repeats until the total resource budget is exhausted.
\end{enumerate}

This iterative, bottleneck-driven approach allows SPARO to adapt the resource allocation dynamically, finding a near-optimal balance between T-state generation and routing capacity tailored to the specific computational demands and error characteristics of the target algorithm, ultimately leading to lower logical error rates compared to static layouts constrained by the same total resource footprint.

\subsection{Generalization to Magic-State Cultivation}
\label{sec:magic_state_cultivation}

SPARO's framework is adaptable to different magic-state preparation methods. While the examples often use distillation~\cite{litinski2019magic}, SPARO can readily incorporate protocols like Magic-State Cultivation~\cite{gidney2024magic}. Cultivation may offer lower spatial overhead but often comes with lower success probabilities, higher sensitivity to physical errors, and dependencies on specific code structures. SPARO treats any preparation method as a probabilistic source defined by its latency, success probability, and output fidelity. This allows SPARO to evaluate the trade-offs involved in using cultivation versus distillation (or a mix) and incorporate the corresponding resource requirements and probabilistic effects into its scheduling and dynamic allocation decisions, maintaining its goal of optimizing overall FTQC performance.

\section{Evaluation}
\subsection{Logical Error Rate}

\begin{figure}[htbp]
    \centering
    \begin{subfigure}[b]{0.98\linewidth}
        \centering
        \includegraphics[width=\linewidth]{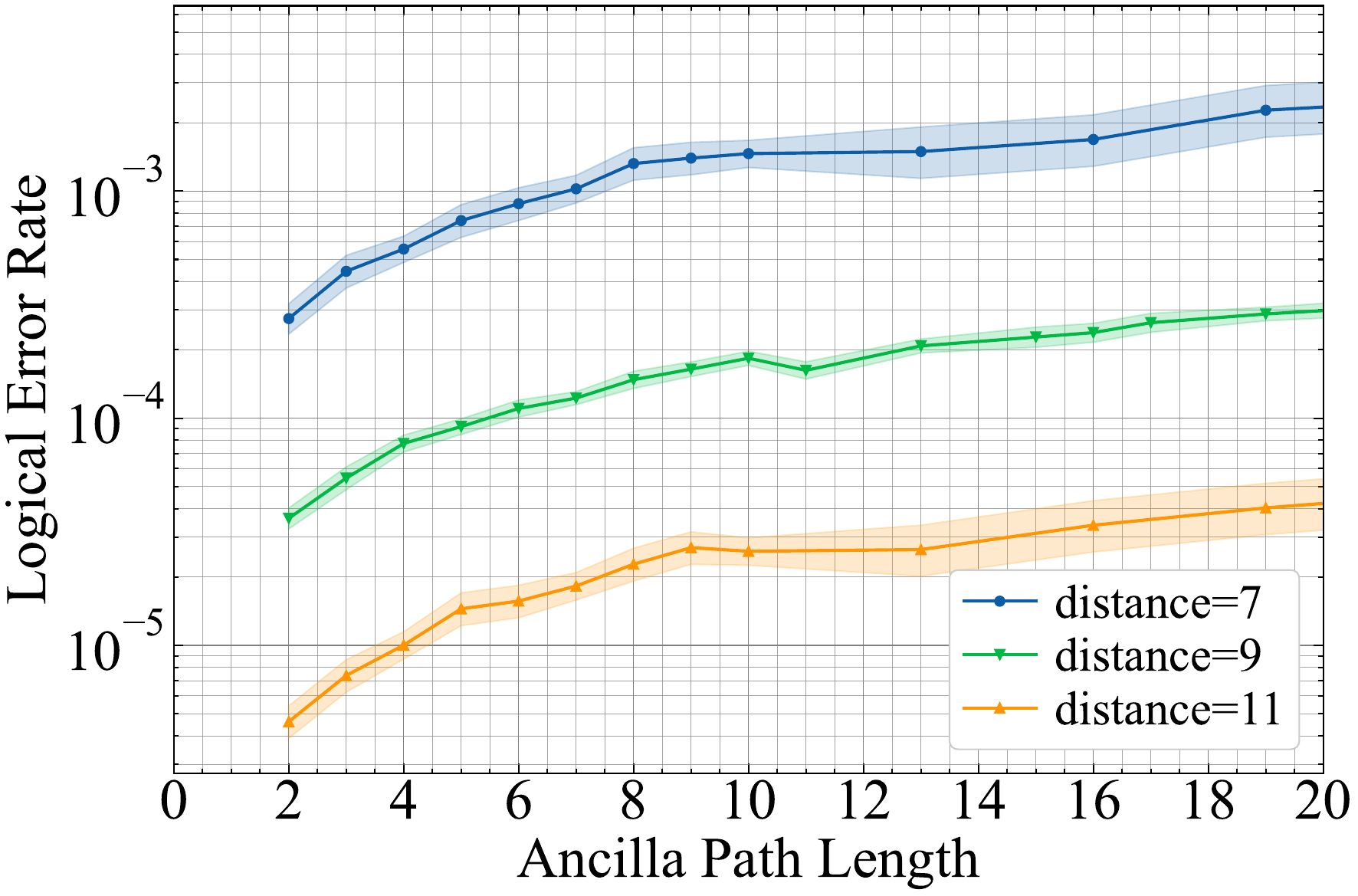}
        \caption{Pauli Product Measurement}
    \end{subfigure}
    \hfill
    \begin{subfigure}[b]{0.98\linewidth}
        \centering
        \includegraphics[width=\linewidth]{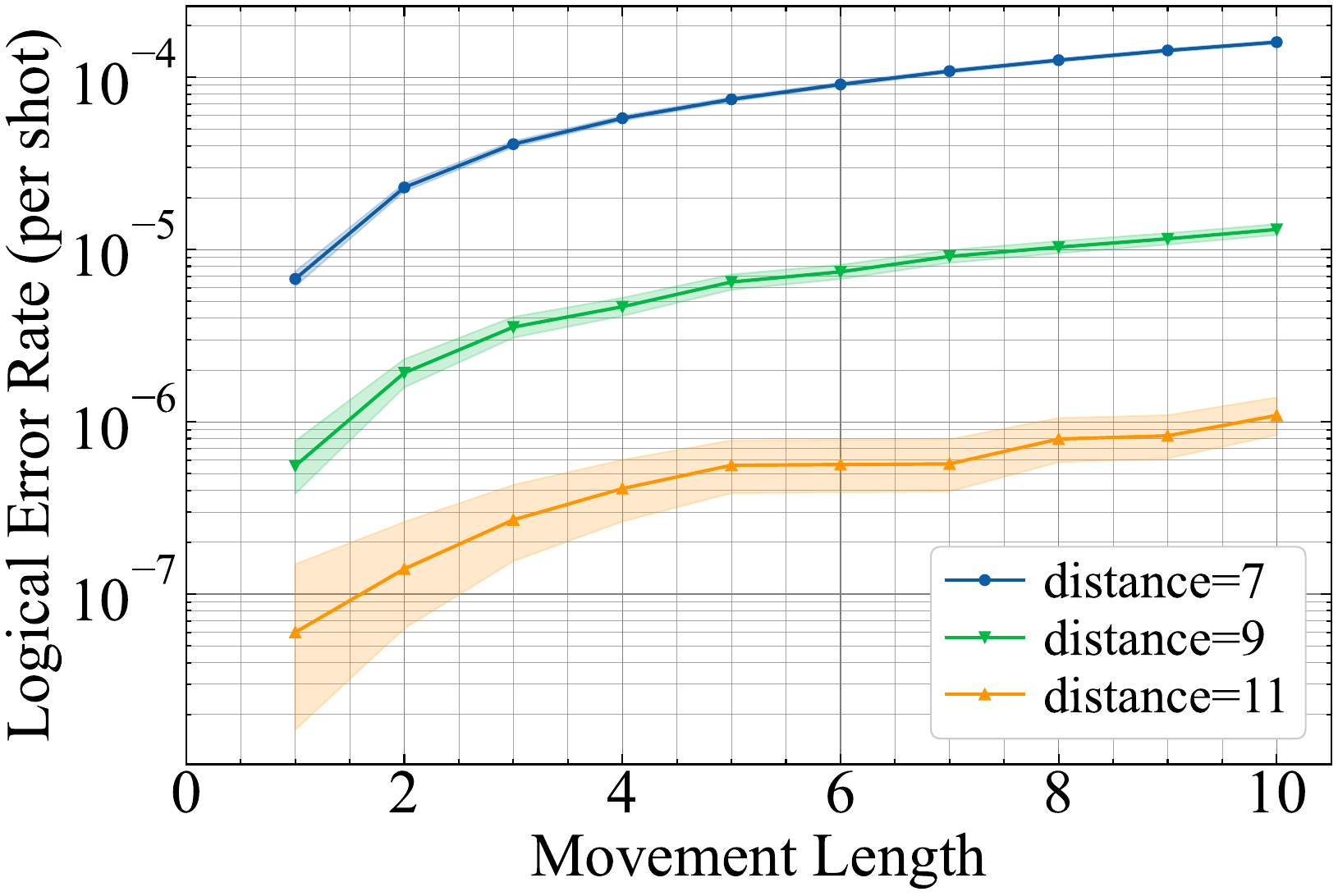}
        \caption{Logical Qubit Movement}
    \end{subfigure}
    \caption{Logical error rates of Pauli product measurements performed in one time step. The X-axis represents the length of the formed ancilla path. The highlighted regions indicate hypotheses whose conditional probabilities \( P(\text{hypothesis} \mid \text{data}) \) are within a factor of 1000 of the most probable hypothesis, corresponding to a Bayes factor \( < 1000 \).}
    \label{fig:ppm}
\end{figure}

We employ Monte Carlo sampling with the STIM package~\cite{gidney2021stim}, a high-performance Clifford stabilizer simulator to model the behavior of large-scale quantum error correction. Our simulations adopt a standard phenomenological noise model with three primary error channels:
\begin{itemize}
    \item \textbf{State Preparation and Measurement (SPAM) Error}:
    This channel models unintended \(X\) or \(Z\) flips after physical qubit initialization or immediately before measurement, depending on the associated basis.

    \item  \textbf{Two-Qubit Depolarizing Error}:  
    During syndrome extraction, which typically employs two-qubit gates (e.g., \(CX\) or \(CZ\)), this channel applies a depolarizing noise model. Specifically, with probability \(p_{\mathrm{dep2}}\!/16\), one of the sixteen two-qubit Pauli errors (\(\{II, IX, IY, IZ, XI, XX, XY, XZ, \ldots, ZZ\}\)) is introduced uniformly.

    \item \textbf{Single-Qubit Depolarizing Error.}:  
    For each one-qubit gate (e.g., a Hadamard or phase gate), this channel applies a depolarizing noise with probability \(p_{1q}\), uniformly selecting one of the three single-qubit Pauli errors \((X, Y, Z)\).

    \item \textbf{Idling Depolarizing Error}:  
    At every time step in which a qubit remains idle (i.e., while other qubits are performing operations), there is a probability \(p_{\mathrm{idle}}\) of incurring one of the three Pauli errors \((X, Y, Z)\).
\end{itemize}

\textbf{Pauli-Product-Measurement Error Scaling}:
Figure~\ref{fig:ppm}(a) shows the logical error rate of a multi-qubit Pauli product measurement under a physical error rate of \(p = 0.001\) for three surface-code distances, \(d = 7\), \(9\), and \(11\). Increasing the code distance by two units reduces the logical error rate by approximately an order of magnitude, demonstrating the characteristic exponential suppression of the surface code. Meanwhile, the horizontal axis indicates the ancilla length, whose growth leads to an almost linear increase in logical error. This effect arises from the extended interaction path required when measuring a larger number of qubits, which introduces more fault opportunities in the ancilla patches. 
% Notably, these results differ markedly from the error behavior of idle qubits at the same distances: multi-qubit PPMs face additional failure modes through simultaneous interactions among multiple qubits, highlighting the operational overhead tied to complex measurements.

\begin{figure*}[htbp]
  \centering
  \begin{subfigure}[b]{0.33\textwidth}
    \centering
    \includegraphics[width=\linewidth]{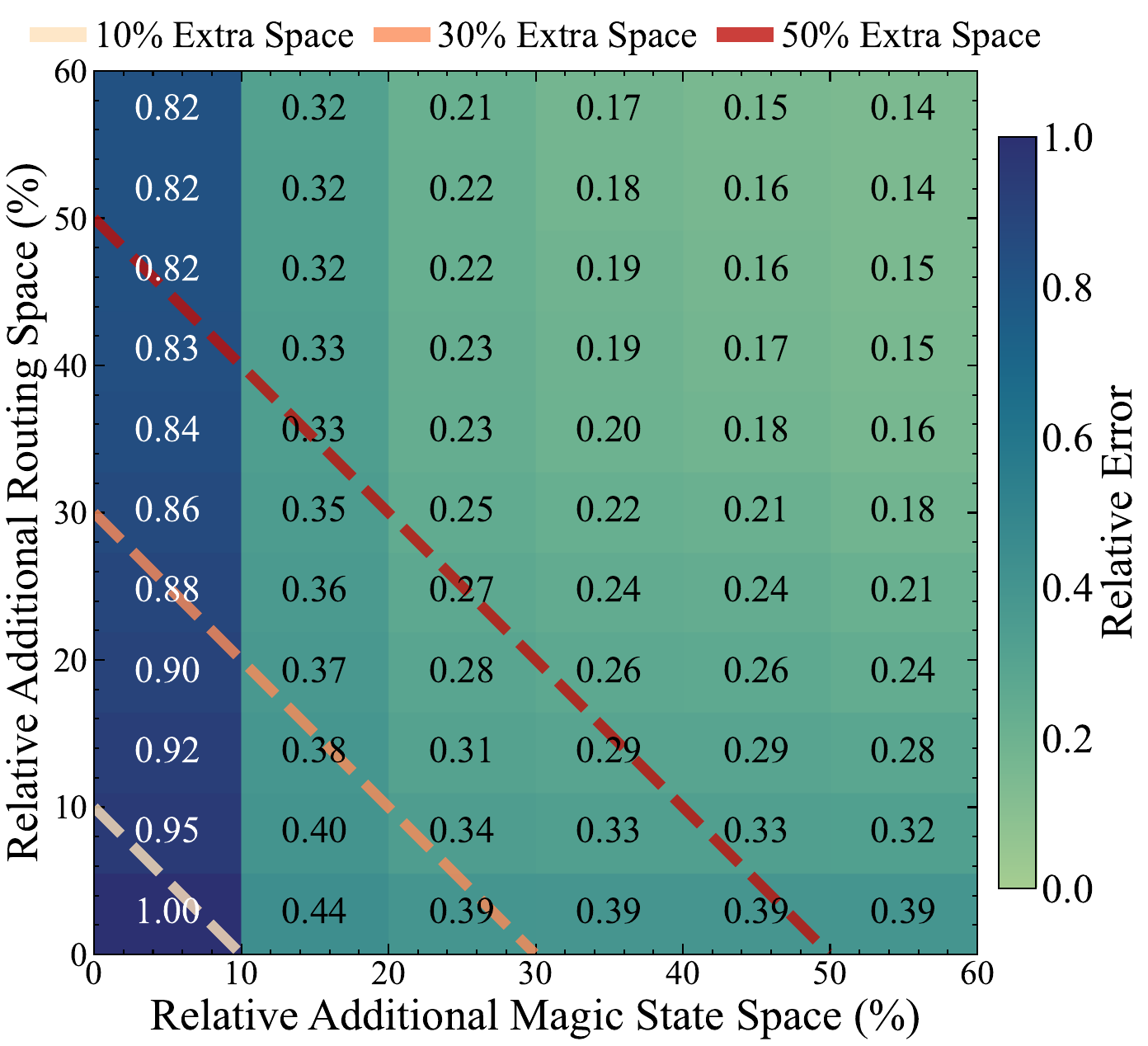}
    \caption{115-qubit Swap Test ($T=680$, $w_{avg}=1.34$).}
    \label{fig:swaptest115}
  \end{subfigure}\hfill
  \begin{subfigure}[b]{0.33\textwidth}
    \centering
    \includegraphics[width=\linewidth]{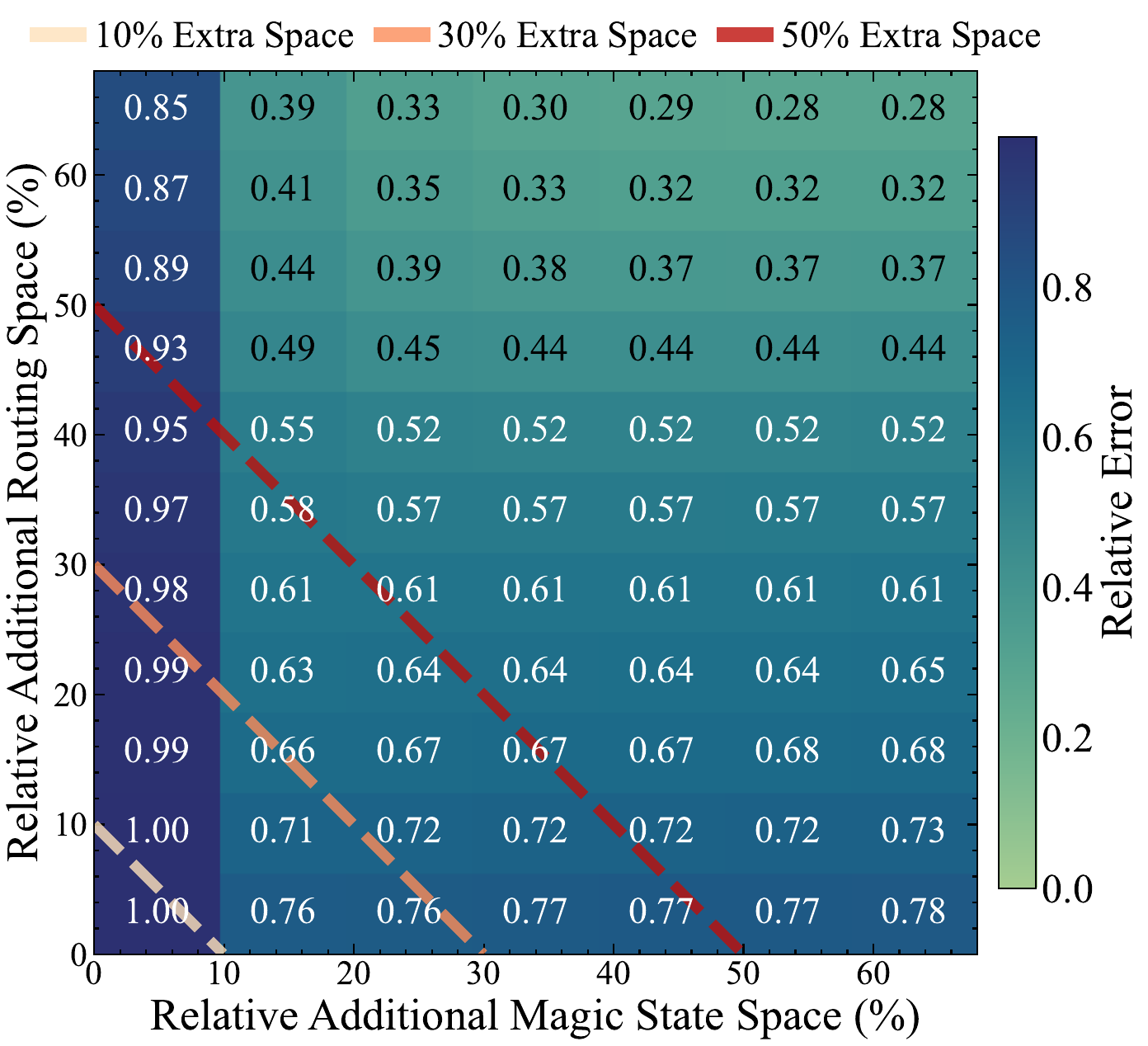}
    \caption{118-qubit Adder ($T=416$, $w_{avg}=33.5$).}
    \label{fig:adder118}
  \end{subfigure}\hfill
  \begin{subfigure}[b]{0.33\textwidth}
    \centering
    \includegraphics[width=\linewidth]{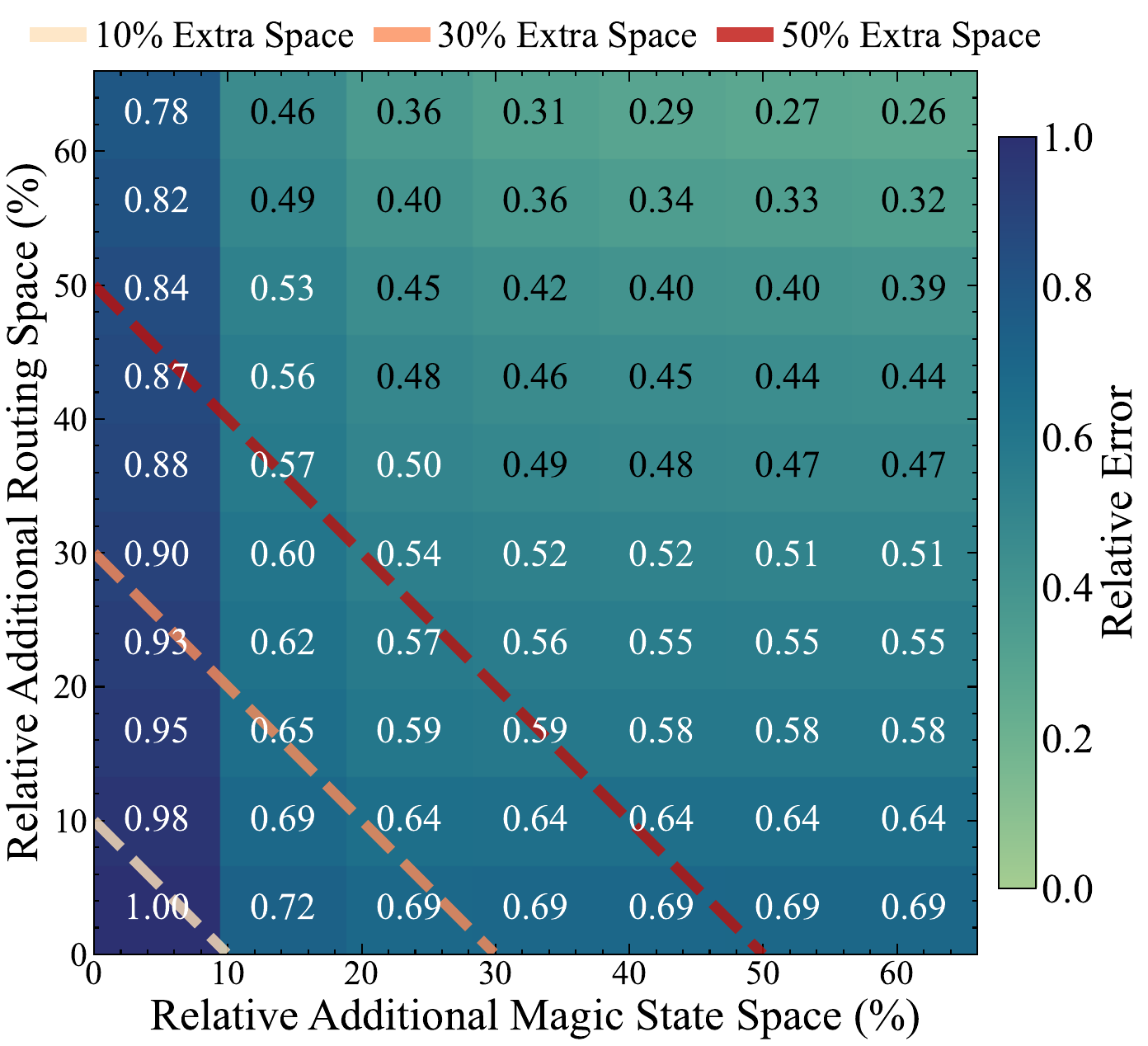}
    \caption{56-qubit QFT ($T=1536$, $w_{avg}=112.25$).}
    \label{fig:qft56}
  \end{subfigure}
\caption{
Heatmaps of relative logical error rates under varying extra-resource allocations for three circuits: (a) 115-qubit Swap Test (T = 680, $w_{avg}$ = 1.34), (b) 118-qubit Adder (T = 416, $w_{avg}$ = 33.5), and (c) 56-qubit QFT (T = 1536, $w_{avg}$ = 112.25). $T$ is the number of magic-state Pauli product measurements (PPMs) and $w_{avg}$ is the average PPM weight. In each subplot, the horizontal axis denotes additional magic state factory resources, the vertical axis denotes additional routing space. These additional resources are expressed relative to a minimal layout baseline for that algorithm. Dashed diagonal lines indicate total extra resource budgets of 10\%, 30\%, and 50\%. Lower heatmap values correspond to lower relative logical error rates and are better.
}
  \label{fig:all-heatmaps}
\end{figure*}

\textbf{Moving Logical Operator Error Scaling}:
Figure~\ref{fig:ppm}(b) presents the logical error rates for moving logical operators by \(n\)~tiles on the surface-code grid, also performed within a single time step. A similar trend to the PPM results emerges, in which increasing the number of moved tiles leads to a higher logical failure rate. Conceptually, the number of tiles moved parallels the ancilla-path length in the PPM setting as both capture how many ancillary patches must be activated by stabilizers, thus determining the net space-time volume of the operation. In Pauli-based computation, such moving operations are closely related to logical-patch , since boundary reorientation often entails a one-tile shift forward and, in cramped layouts, another shift back to the original location. Consequently, the error model for moving operators provides a practical estimate of the logical failures incurred by boundary reconfiguration and short-range patch motion required for certain rotations.

In the following logical error rate results, we adopt a distance-9 surface code with a physical error rate of 0.001. The minimal layout is used as the baseline for benchmarking the logical error rate, and all reported error rates are given as relative values with respect to this baseline.

\subsection{Heatmap of Resource Allocation for Quantum Algorithm}

Fig~\ref{fig:all-heatmaps} shows the heatmap of relative error rate for 3 circuit: 115-qubit Swap Test (T = 680, w = 1.34).1 18-qubit Adder (T = 416, w = 33.5) c) 56-qubit QFT (T = 1536, w = 112.25). with additional resource allocation to magic state space and routing region. 

 Across all three heatmaps, the marginal benefit in logical error‐rate reduction decreases as resources are added to only magic state factories or routing regions. Initially, increasing $T$-factory capacity by up to approximately 20–25\% yields a substantial decline in error rate (56\% for swap test, 24\% for adder and 28\% for qft )by reducing magic‐state idle time. Beyond this point, additional factory capacity remains underutilized and produces no fidelity gains, as the primary bottleneck shifts to rotations and high-weight Pauli‐product operations. A similar trend holds for routing space: the marginal  reductions in logical error rate diminish when additional routing spaces reaches certain number depending on the additional magic tsate space. These results demonstrate that, under a fixed total hardware budget, the optimal strategy is to allocate extra resources in a balanced manner between $T$-factory capacity and routing area, thereby achieving the greatest overall reduction in logical errors.

Comparing the three circuits, the 115-qubit Swap Test exhibits the greatest benefit from additional resources, achieving a 73\% reduction in logical error rate with 50\% extra hardware, compared to 39\% for the 118-qubit Adder and 43\% for the 56-qubit QFT.  This difference stems from the Swap Test’s low degree of entanglement. Its average Pauli-product weight is only 1.34, versus 33.5 and 1.74 for the Adder and QFT, respectively and the relatively low number of $\pi/8$ rotations it performs.  In contrast, the more highly entangled Adder and QFT remain bottlenecked by high weighy pauli product operations and rotation demands even when given extra space.  These findings underscore that, under a fixed resource budget, the optimal allocation between $T$-state factories and routing area must be tailored to each algorithm’s entangling structure to minimize logical errors.

% This heatmap comparison also provides valuable insights into the performance optimization potential for executing quantum algorithms in fault-tolerant settings. Notably, highly entangled circuits generally require a greater allocation of resources to achieve performance levels comparable to those of less complex circuits. In such cases, additional space for both $T$-state factories and routing is essential in order to mitigate the overheads associated with inter-qubit communication and complex measurement operations, thereby maintaining an acceptable logical error rate.

\begin{figure}[!ht]
    \centering
    \includegraphics[width=1\linewidth]{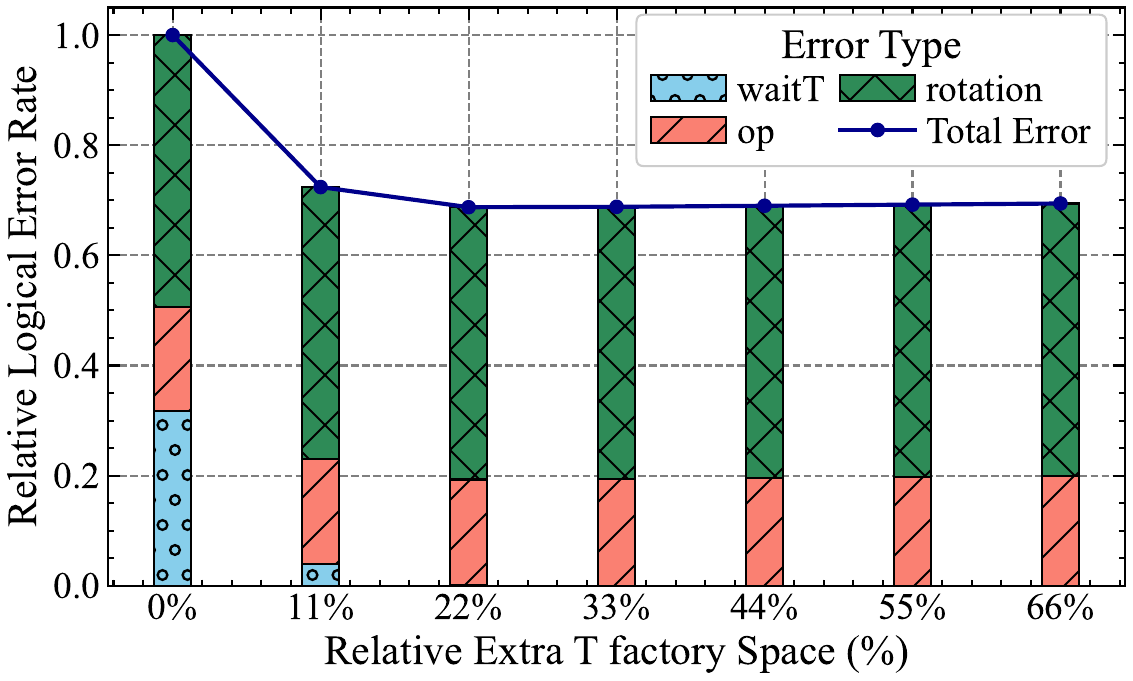}
    \caption{Breakdown of error source for 56-qubit QFT circuit when only adding additional resources for magic states. }
    \label{fig:terror}
\end{figure}

\begin{figure*}[htbp] % Use placement specifier like [htbp]
    \centering % Center all 
    \begin{subfigure}{0.33\linewidth} % Adjusted width for three figures in a row
        \centering
        \includegraphics[width=\linewidth]{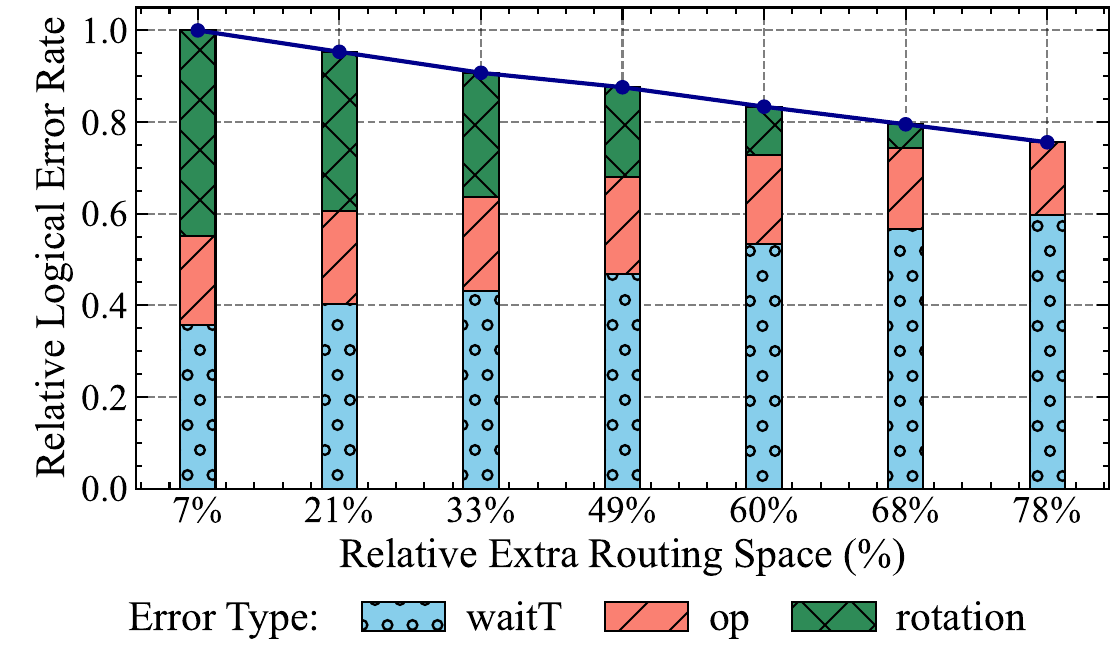}
        \caption{Routing with 1 $T$-factory.} % Simplified caption
        \label{fig:overhead_1T_routing} % Specific label
    \end{subfigure}
    \hfill % Add horizontal space
    \begin{subfigure}{0.33\linewidth}
        \centering
        \includegraphics[width=\linewidth]{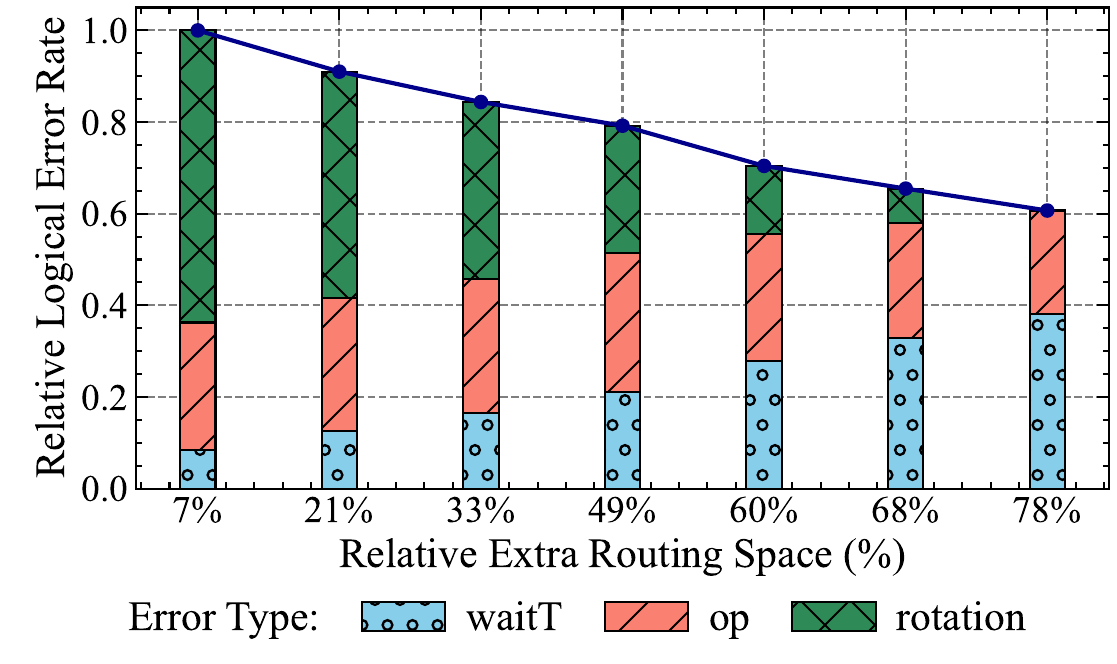}
        \caption{Routing with 2 $T$-factories.} % Simplified caption
        \label{fig:overhead_2T_routing} % Specific label
    \end{subfigure}
    \hfill % Add horizontal space
    \begin{subfigure}{0.33\linewidth}
        \centering
        \includegraphics[width=\linewidth]{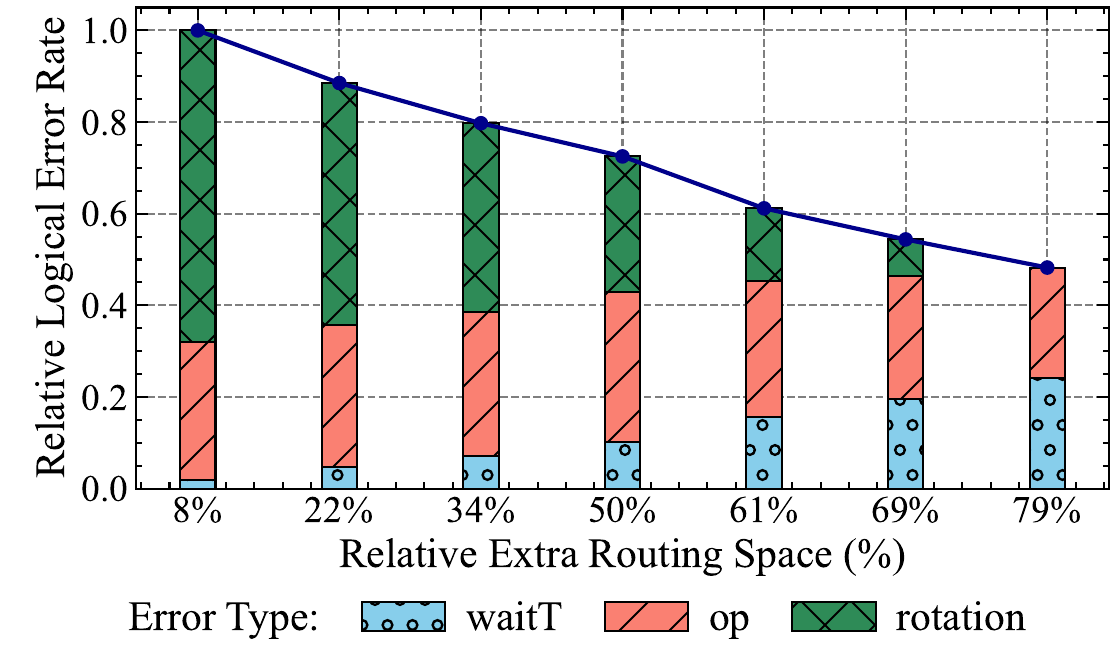}
        \caption{Routing with 3 $T$-factories.} % Simplified caption
        \label{fig:overhead_3T_routing} % Specific label
    \end{subfigure}

    \caption{\textbf{Breakdown of error source for 56-qubit QFT when adding routing capacity alongside $T$-factories.} The overlaid blue line plots the total normalized error. Increasing routing space reduces rotation-induced errors but reveals $T$-generation latency as the dominant error; adding extra $T$-factories reduced the overall errors.}
\label{fig:overhead_routing_tfactories} % Main label for this figure group
\end{figure*}

\subsection{Error Breakdown for Resource Allocation for Magic State and Routing Region}

In Fig.~\ref{fig:terror} and \ref{fig:overhead_routing_tfactories}, we break down the logical error rate for 56-qubit QFT circuit from \ref{fig:qft56} into three categories to help understand the major error source: 
\begin{enumerate}
    \item \textbf{Operations} (red): The multi-qubit Pauli product measurements. It is primarily affected by the ancilla path length. 
    \item \textbf{Rotations} (green): Time spent adjusting logical-qubit boundaries to enable certain measurements or gate decompositions.
    \item \textbf{Wait for \emph{T}} (blue): Idle time awaiting magic-state generation.
\end{enumerate}

Notably, rotation steps and $T$-wait times often interleave: when rotation was previously the bottleneck, $T$-generation latency could be masked. Once we add more routing resources, rotation delays diminish, unmasking the $T$-wait overhead. Consequently, the proportion of time (and error contribution) attributed to $T$-waiting becomes larger, revealing that magic-state generation can then become the dominating factor. 

\textbf{Comparisons Across Different Magic State Allocations}:
Fig.~\ref{fig:terror} shows the effect of adding additional $T$-state factories, illustrating that increased availability of $T$-states reduces the waiting overhead (\texttt{waitT}). The first additional magic-state factory lead to a marked reduction in the overall error rate, primarily due to the reduction in idle time required for magic-state generation. However, beyond this point, the improvements diminish because the large rotation cost remains unmitigated. In such cases, further increases in $T$-state availability result in a scenario where some magic-state idling continues concurrently with rotations, as the error contribution from waiting for $T$-states is already minimal.

\textbf{Comparisons Across Different Routing Region Allocations}:
Fig.~\ref{fig:overhead_routing_tfactories} presents three subplots illustrating how the relative logical error rate changes as additional resources are allocated to routing, given a fixed number of $T$-factories. Fig \ref{fig:overhead_1T_routing} assumes a single $T$-factory, while Fig \ref{fig:overhead_2T_routing} and Fig \ref{fig:overhead_3T_routing} consider two and three factories. In each subplot, the $x$-axis indicates the percentage of extra routing space, and the $y$-axis shows the resulting logical error rate normalized by a minimal-layout baseline.  

A clear trend emerges: as the routing allocation increases from left to right, the \texttt{rotation} overhead (green) generally decreases because there is more space to maneuver logical qubits and ancillas, resulting in fewer idle patches waiting for reconfiguration. At the same time, if only one $T$-factory is available (subfigure~(a)), the \texttt{waitT} segment can remain significant despite added routing, reflecting a bottleneck in magic-state generation. In subfigure~(b), with two factories, this waiting overhead is reduced, and routing becomes a more critical factor in lowering the total error. By subfigure~(c), where three $T$-factories minimize waiting, any further improvements depend primarily on reducing rotation overhead via expanded routing. 

This result provides a in-depth view from the heatmap comparison in understanding how resource allocation shifts the bottleneck under PBC compute models. Figure~\ref{fig:all-heatmaps},\ref{fig:terror} and \ref{fig:overhead_routing_tfactories} collectively underscore that simply adding one type of resource (e.g., more $T$-factories or a larger routing region) can shift, rather than reducing, the overall bottleneck. As a result, achieving the most effective performance improvement may require a balanced allocation of space between $T$-factory capacity and routing flexibility.

\subsection{\sol~ benchmarked against compact and intermediate block}

\begin{figure*}[ht]
    \centering
    \includegraphics[width=1\textwidth]{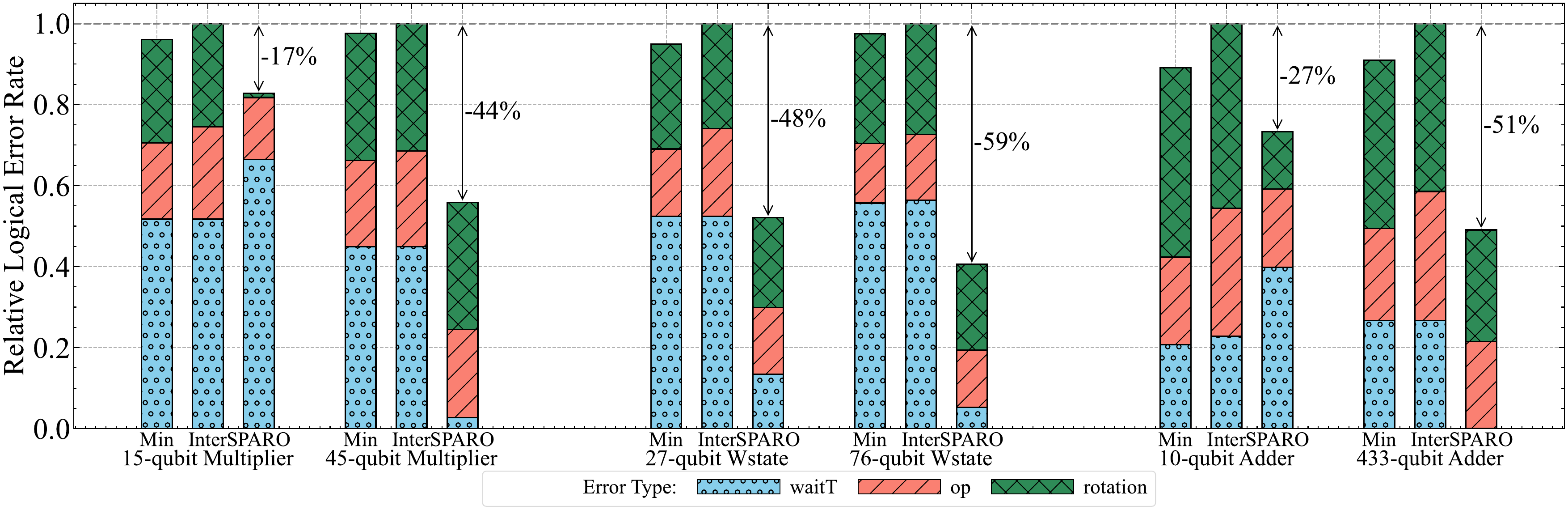}
    \caption{Comparison of logical error rate from the Minimal, Intermediate, and \sol~layout strategies. The Minimal layout uses the fewest space. The Intermediate layout from ~\cite{litinski2019game} allocates about 30\% more space.  The \sol~ approach matches the Intermediate layout’s total qubit budget but dynamically allocates those resources per algorithm. Across all benchmarks, \sol{} consistently attains greater reductions in logical error rate for larger instances of the same circuit type compared to the Intermediate layout.}
    \label{fig:layout_comparison}
\end{figure*}

Figure~\ref{fig:layout_comparison} compares the performance of SPARO against the defined Minimal and Intermediate static layouts for several benchmark algorithms. The Intermediate layout uses approximately 30\% more tiles than the Minimal baseline. For this comparison, the SPARO configurations shown (\textit{SPARO}) are optimized using a total resource budget \textbf{identical to the number of tiles used by the Intermediate layout}. The bars show the breakdown of logical error contributions (WaitT, Op, Rotation) for each layout strategy.

The \textbf{Minimal} layout uses the fewest tiles and is the most space-efficient solution. However, with only one magic‐state factory and minimal routing, the Minimal layout incurs excessive idle waits and costly patch rotations, yielding higher logical error rates.

The \textbf{Intermediate} layout as proposed in \cite{litinski2019game} and employs roughly 30\% more tiles than the Minimal layout. The additional space helps alleviate routing congestion and reduces the rotations with larger routing space. Despite these improvements, the Intermediate layout remains a static configuration and may not be optimal for every quantum algorithm.

In contrast, the \textbf{\sol} layout employs a dynamic strategy. Although it uses the same total number of tiles as the Intermediate layout, it reconfigures the tile allocation dynamically for each algorithm based on its specific needs. As a result, the SOL layout typically achieves lower overall logical error rate by effectively balancing resource constraints related to routing, magic-state factories, and patch rotations.

Figure~\ref{fig:layout_comparison} shows that, with the same footprint, \sol’s dynamic allocation consistently outperforms the static Intermediate layout, especially as circuits scale. For example, \sol achieves a $44\%$ error reduction on the 45-qubit multiplier versus only $17\%$ on the 15-qubit version; $59\%$ versus $48\%$ on the 76-qubit versus 27-qubit W-state; and $51\%$ versus $27\%$ on the 433-qubit versus 10-qubit adder.

% \begin{figure}
%     \centering
%     \includegraphics[width=0.95\linewidth]{image.png}
%     \caption{Caption}
%     \label{fig:enter-label}
% \end{figure}
% This figures shows intermediate block, minmimal block and sol selected layout comparsion. The minimal layout has the fewest titles in the total grid. While the intermediate is as suggested in game of surface code that has about 30 percent higher space compared to minimal and \sol is to use sol to dynamically select layout for each algorithm using ths same number of tiles as intermediate result. 
% The result shows that 

\subsection{Scalability}
Table~\ref{tab:runtime-pauli-stats-compact} reports the wall‐clock time required by \sol{} to compile and optimize a set of PBC‐transpiled quantum circuits, with logical qubit counts ranging from $n=8$ to $n=433$.  For each circuit we also list its total number of Pauli product measurement with magic state as $T$ and the Pauli‐product measurement weight statistics( [min, avg, max]) with the standard deviation.  These quantities are the primary determinants of \sol{}’s runtime performance.  All experiments were executed on a MacBook Pro equipped with an Apple M3 Max chip. 

Notably, even at 433~qubits (\texttt{adder\_n433}), \sol{} completes the scheduling and optimizations in approximately 10 second. For mid-sized circuits (tens to hundreds of qubits), runtime overheads generally fall below a few seconds, reflecting the \sol’s efficient handling of state-space transformations and classical optimizations for qubit mapping, Steiner-tree routing for Pauli product measurements, and rotation scheduling. 

In practice, these compilation or scheduling steps occur offline and well before actual quantum hardware execution. Therefore, the observed sub-minute runtimes is acceptable for near-term devices. As circuit sizes continue to increase, \sol{}’s underlying heuristics can scale to hundreds or potentially thousands of qubits without incurring prohibitive slowdown. 

Additionally, each tile in our layout represents a distance-$d$ surface-code patch containing $2d^2 - 1$ physical qubits (e.g., 161 qubits at distance 9). A layout of 100 such tiles at distance 9 thus includes 16,100 physical qubits—significantly exceeding the scale of current quantum devices. And for 433-qubit adder circuit, the system contains more than 10 million physical qubits.

Consequently, the data in Table~\ref{tab:runtime-pauli-stats-compact} demonstrate that \sol{} is capable of managing realistic, larger-than-near-term quantum workloads within practical timescales, underscoring its applicability as a co-design tool for future fault-tolerant quantum machines and algorithms.
\begin{table}[ht]
\centering
\caption{Wall‐clock runtimes, total magic state Pauli product measurement(T), and their weight statistics ([min, average, max]) with standard deviation for various circuits, ordered by logical qubit count \(n\).}
\label{tab:runtime-pauli-stats-compact}
\begin{tabular}{@{}l  c  r  l  r@{}}
\toprule
\textbf{Circuit}      & \textbf{Time (s)} & \textbf{T}  & \textbf{Weight} & \textbf{Std dev} \\
\midrule
dnn\_n8           & 1.92    & 2960   & [1, 3.42, 8]    & 2.13   \\
qpe\_n9           & 0.055  & 86     & [1, 1.70, 4]    & 0.55   \\
seca\_n11         & 0.0087 & 16     & [2, 3.25, 6]    & 1.24   \\
multiplier\_n15   & 0.034  & 42     & [1, 3.81, 6]    & 1.25   \\
bigadder\_n18     & 0.045  & 32     & [3, 6.62, 11]   & 2.37   \\
qram\_n20         & 0.025  & 48     & [1, 2.54, 7]    & 1.76   \\
ising\_n26        & 0.24   & 324    & [1, 1.25, 3]    & 0.47   \\
multiplier\_n45   & 0.68   & 396    & [1, 4.38, 6]    & 1.17   \\
vqe\_uccsd\_n8    & 3.08    & 3742   & [1, 5.35, 8]    & 1.49   \\
ising\_n10        & 0.42   & 928    & [1, 1.30, 3]    & 0.54   \\
multiply\_n13     & 0.0055 & 14     & [1, 1.86, 3]    & 1.03   \\
qf21\_n15         & 0.13   & 162    & [1, 1.69, 4]    & 0.54   \\
qft\_n18          & 1.40    & 1333   & [1, 4.30, 11]   & 3.09   \\
swap\_test\_n25   & 0.11   & 140    & [1, 1.34, 3]    & 0.63   \\
wstate\_n27       & 0.38   & 416    & [1, 2.04, 11]   & 2.26   \\
qft\_n56          & 7.46    & 2911   & [1, 11.74, 33]   & 10.51   \\
swap\_test\_n115  & 1.59    & 680    & [1, 1.34, 3]    & 0.63   \\
adder\_n118       & 2.54    & 416   & [3, 33.5, 118]& 36.12 \\
adder\_n433       & 10.43    & 1536   & [3, 112.25, 433]& 137.74 \\
\bottomrule
\end{tabular}
\end{table}

\section{Related Work}
Several prior works have explored transpiling logical circuits onto surface‐code architectures using braiding rather than lattice surgery. Ding et al. \cite{ding2018magic} were the first to propose hardware functional units for magic‐state factories within the braid framework. More recently, Hua et al. introduced Autobraid \cite{hua2021autobraid}, a tool that automates defect‐braid mapping and routing of arbitrary logical operations onto the surface code. In the context of lattice surgery, most prior studies on compiling quantum algorithms have focused on Clifford+T circuits \cite{Watkins2024highperformance,leblond2023tiscc}. Watkins et al.\cite{Watkins2024highperformance} proposed a high‐performance compiler for Clifford+T, achieving rapid compilation speeds but with suboptimal qubit mapping and rerouting. Leblond et al.\cite{leblond2023tiscc} presented a two‐stage protocol that first converts Clifford+T circuits into a lattice‐surgery intermediate representation and then assigns operations to specific hardware tiles. More recently, Tan et al.’s LaSsynth \cite{Tan2024SAT} synthesizes lattice‐surgery subroutines via a SAT‐based model. A recent work by Silva et al. \cite{silva2025optimizingmultilevelmagicstate} introduces a multi‐level magic-state factory architecture that optimizes both the number of distillation levels and their code distances to minimize space–time overhead under a user‐defined error budget. This can be seamlessly integrated with \sol’s dynamic resource‐allocation framework, allowing both routing capacity and factory throughput to be co‐optimized for any target quantum algorithm.
\section{Discussion}
\subsection{Generality to other Quantum Error Correction Code}
In this work, we focus on the surface code computational model due to its well-understood structure and practical feasibility for near-term quantum devices\cite{google2023suppressing,acharya2024quantum}. While alternative quantum error-correcting codes, such as high-yield quantum low-density parity-check (qLDPC) codes, are promising candidates for improving encoding rates~\cite{cross2024linear, bravyi2024high}, they typically involve more complex topological layouts and currently lack clear, scalable paths to universal fault-tolerant computation. Although our proposed method, \sol, is developed within the surface code framework, its core insights into resource allocation bottlenecks and algorithmic performance can be generalized to other quantum error correction codes. Notably, in qLDPC codes such as the Gross code, Pauli product measurements often incur higher costs, shifting the optimization priorities. However, such changes in the error and cost models are naturally accounted for by \sol.

\section{Conclusion}

Achieving practical fault-tolerant quantum computing over surface code architectures necessitates optimizing the significant resource overheads inherent in managing computation, state distillation, and routing. Static architectural designs often struggle to adapt to the varying demands of different quantum algorithms, leading to performance bottlenecks and inflated error rates. This work introduced SPARO, a comprehensive framework designed to bridge this gap by enabling dynamic, algorithm-aware resource allocation for surface-code architectures implementing PBC via lattice surgery.

SPARO integrates a detailed logical error model, accounting for Pauli product measurements, qubit idling, and patch manipulation overheads, with an end-to-end compilation pipeline. By analyzing the specific patterns and computational bottlenecks of a given algorithm, identifying trade-offs between routing availability and magic-state generation throughput, SPARO adaptively partitions hardware resources. Our simulations demonstrate that this dynamic approach can significantly reduce logical error rates compared to fixed, static layouts, by tailoring the architecture (e.g., expanding routing regions or adding magic-state factories) to meet the immediate needs of the computation.

The results underscore the importance of considering active computation dynamics and adopting flexible, co-design strategies for optimizing fault-tolerant systems. SPARO provides a framework for exploring these architectural trade-offs, demonstrating that algorithm-specific architectures identified through its architectural optimization search can achieve lower logical error rates within a given resource budget compared to established static block designs, such as those proposed in Litinski's "Game of Surface Codes"~\cite{litinski2019game}, even when allocated the identical total hardware resource footprint. \sol~will be made available as open-source software to facilitate further research in algorithm-architecture co-optimization for fault-tolerant quantum computing.

\section*{Acknowledgements}

This research was supported by PNNL’s Quantum Algorithms and Architecture for Domain Science (QuAADS) Laboratory Directed Research and Development (LDRD) Initiative. The Pacific Northwest National Laboratory is operated by Battelle for the U.S. Department of Energy under Contract DE-AC05-76RL01830. This research used resources of the Oak Ridge Leadership Computing Facility, which is a DOE Office of Science User Facility supported under Contract DE-AC05-00OR22725. This research was also supported in part by the National Science Foundation (NSF) under grant agreements 2329020 and 2301884.  

% \begin{figure}
%     \centering
%     \includegraphics[width=1\linewidth]{Screenshot 2025-04-09 at 7.47.50 AM.png}
%     \caption{Trade offs}
%     \label{fig:enter-label}
% \end{figure}
% \begin{figure}
%     \centering
%     \includegraphics[width=1\linewidth]{Screenshot 2025-04-10 at 6.44.10 AM.png}
%     \caption{Trade offs}
%     \label{fig:enter-label}
% \end{figure}

%%%%%%% -- PAPER CONTENT ENDS -- %%%%%%%%

\bibliographystyle{ACM-Reference-Format}
\bibliography{refs}

\end{document}